\documentclass[12pt]{article}
\pdfoutput = 1

\usepackage{xcolor}
\usepackage[utf8]{inputenc} % allow utf-8 input
\usepackage[T1]{fontenc}    % use 8-bit T1 fonts
\usepackage{hyperref}       % hyperlinks
\usepackage{url}            % simple URL typesetting
\usepackage{booktabs}       % professional-quality tables
\usepackage{amsfonts}       % blackboard math symbols
\usepackage{nicefrac}       % compact symbols for 1/2, etc.
\usepackage{microtype}      % microtypography
\usepackage{lipsum}
\usepackage{moreverb}
\usepackage{subfig}
\usepackage{afterpage}
\usepackage{listings}
\usepackage{algorithmic}
\usepackage{amsmath,amssymb,amsfonts}
\usepackage{amsthm}
\usepackage{multirow}
\usepackage{hyperref}
\usepackage{mathtools}
\usepackage{bm}
\usepackage{mathrsfs}
\usepackage{rotating}

\usepackage{diagbox}

\usepackage{graphicx}
\usepackage{epstopdf}

\usepackage{natbib}
\setcitestyle{aysep={}}

\DeclareMathOperator{\Var}{\operatorname{Var}}
\DeclareMathOperator{\Cov}{\operatorname{Cov}}
\DeclareMathOperator{\Corr}{\operatorname{Corr}}
\DeclareMathOperator{\E}{\mathbb{E}}

\newcommand{\blind}{0}
\newcommand{\newtextforblind}[2]{\if0#1{#2}\fi
\if1#1{\textcolor{blue}{#2}}\fi}
\begin{document}

\def\spacingset#1{\renewcommand{\baselinestretch}%
{#1}\small\normalsize} \spacingset{1}

%%%%%%%%%%%%%%%%%%%%%%%%%%%%%%%%%%%%%%%%%%%%%%%%%%%%%%%%%%%%%%%%%%%%%%%%%%%%%%

\if0\blind
{
  \title{\bf Profile Monitoring via Eigenvector Perturbation}
  \date{November 6, 2024}
  \author{Takayuki Iguchi\thanks{
    T. Iguchi is Assistant Professor, Air Force Institute of Technology. The work was part of his PhD dissertation at Florida State University. 
    This research was supported by the Office of the Secretary of Defense, Directorate of Operational Test and Evaluation under Grant FA8075-14-D-0019 and the Test Resource Management Center under the Science of Test research program grant FA807518F1525.
    The views expressed in this article are those of the author and do not reflect the official policy or position of the United States Air Force, Department of Defense, or the U.S. Government.
    This article has been accepted for publication in {\it Technometrics}, published by Taylor \& Francis.
    }\hspace{.2cm}\\
    Department of Statistics, Florida State University\\
    and \\
    Andr\'{e}s F. Barrientos \hspace{.2cm}\\
    Department of Statistics, Florida State University\\
    and \\
    Eric Chicken \hspace{.2cm}\\
    Department of Statistics, Florida State University\\
    and \\
    Debajyoti Sinha \hspace{.2cm}\\
    Department of Statistics, Florida State University}
  \maketitle
} \fi

\if1\blind
{
  \bigskip
  \bigskip
  \bigskip
  \begin{center}
    {\LARGE\bf Profile Monitoring via Eigenvector Perturbation}
\end{center}
  \medskip
} \fi

\bigskip
\begin{abstract}
\noindent In Statistical Process Control, control charts are often used to detect undesirable behavior of sequentially observed quality characteristics. 
Designing a control chart with desirably low 
False Alarm Rate  (FAR) and detection delay ($ARL_1$) is an important challenge especially when the sampling rate is high and the control chart has an 
In-Control Average Run Length, called $ARL_0$, of 200 or more, as commonly found in practice.
Unfortunately, arbitrary reduction of the FAR typically increases the $ARL_1$. 
Motivated by eigenvector perturbation theory, we propose the Eigenvector Perturbation Control Chart for computationally fast nonparametric profile monitoring. 
Our simulation studies show that it outperforms the competition and achieves both $ARL_1 \approx 1$ and $ARL_0 > 10^6$. 
\end{abstract}

\noindent%
\vfill

\newpage
\spacingset{1} 
%-----------
%-----------
\section{Introduction}\label{sec:intro}
The main objective of statistical process control (SPC) is to monitor whether a process of interest generating observed quality characteristics is changing over time. 
Our particular SPC context focuses on
designing methods to monitor profiles of quality characteristics.
These profile monitoring methods have found applications in manufacturing processes, crime monitoring, and monitoring objects' locations via wireless technology. For examples, see the review in \cite*
{Maleki2018} and the references therein.
The assumptions about the stochastic models generating the profiles are pivotal in the development of the SPC monitoring methods, usually called control charts.
Increasing complexities of 
processes of interest due to rapid technological developments and automated data acquisition drive
the need for more widely applicable and easily implementable control charts. 
Control charts find use
in diverse fields, including public health surveillance \citep{Bersimis2022PublicHealthMSPC},
stamping operation force monitoring \citep*{Jin1999}, 
artificial sweetener manufacturing \citep{Kang2000}, 
mechanical component manufacturing \citep*{Colosimo2008}, 
and automobile engineering \citep*{amiri2009}. 

Two main challenges exist in control chart design: lowering the False Alarm Rate (FAR) when the underlying process is in-control (IC) (i.e., desirable), and reducing the detection delay after the underlying process becomes out-of-control (OOC) (i.e., undesirable). 
Negative consequences of a high number of false alarms include an increased cost due to unwarranted disruption caused by each false alarm  and a loss of user's confidence in alarms.
For example, in hospital intensive care units, the FAR has detrimental effects on patient safety and effectiveness of care \citep{ImhoffKuhls2006}.
These problems are only exacerbated as the time interval between two consecutive monitoring times decreases, a frequent event with streaming data \citep{WoodallFaltin2019}.
Often, control charts are calibrated to ensure that an average number of monitoring times needed to observe the next false alarm under an IC process, called the IC Average Run Length ($ARL_0$), 
is 200. 
If a process is monitored 
each millisecond, the average time to a false alarm under this level of calibration of $ARL_0$ would be 200 milliseconds,  which is clearly undesirable. 
Widening of control limits may increase the $ARL_0$, but typically comes at the cost of delayed detection of an OOC process. 

In this paper, our goal is to provide an easily implementable  nonparametric control chart applicable 
to a wide variety of quality characteristics, when 
the observed quality characteristics $\bm{y}^t=(y^t_1,
\cdots, y^t_n)$ at  discrete 
monitoring time-point $t$ are noisy observations 
of the scalar function $f^t:\mathbb{R}^d\rightarrow \mathbb{R}$ evaluated at $n$ sampled values $\bm{x}^t_1,
\cdots \bm{x}^t_n$ of a random vector of $d$ explanatory variables. 
So, the observed $y^t_i$ of each functional profile 
has mean $\E[Y^t_i\vert \bm{x}^t_i] = f^t(\bm{x}^t_i)$.
Our goal is to achieve a low FAR and a high $ARL_0$ without sacrificing low detection delay. 
To achieve these goals, our method leverages the $(w\times w)$ sample correlation matrix $\bm{R}$ of the profiles $\bm{y}^{t-w+1}, \dots, \bm{y}^t$ across $w>1$ monitoring times 
(including the current as well as past $w-1$), 
with the $(s,r)$-th entry of $\bm{R}$ being $R_{s,r}=\sum_{i=1}^n (y_i^r - \bar{y}^r)(y_i^s - \bar{y}^s) / \sqrt{\sum_{i=1}^n (y_i^r - \bar{y}^r)^2 \sum_{j=1}^n (y_j^s - \bar{y}^s)^2}$ 
for $(t-w+1)\le s, r\le t$, 
where $\bar{y}^r = \frac{1}{n}\sum_{i=1}^n y_i^r$. 
Essentially, $\bm{R}$ is a consistent estimator of the 
the population correlation matrix $\mathbf{\Gamma}$ whose $(r,s)$-th element is
derived from $\Cov[Y^r, Y^s] = \E[\Cov[Y^r, Y^s \mid \bm{x}^r, \bm{x}^s]] + \Cov[\E[Y^r\mid \bm{x}^r], \E[Y^s \mid \bm{x}^s]]$, where the covariances and expectations are computed using the conditional 
distribution of $ (Y^r, Y^s) \mid \bm{x}^r, \bm{x}^s$ and then the marginal distribution of $(\bm{x}^r, \bm{x}^s) $.
For the sake of brevity, our notations suppress the obvious dependence of $\bm{R}$ on $t$ and dimension $w$.
To propose our control chart, we use the {\it structural} difference in $\bm{R}$ computed from IC profiles when $f^{t-w+1}=\cdots=f^t$ and the $\bm{R}$ computed from mixture of IC and OOC profiles when $f^{k} \ne f^{k+1}$ for some $t-w<k<t$.
Our proposed method
exploits two ideas:
\begin{itemize}
    \item 
    A structural difference exists in $\bm{\Gamma}$ between the case when all $w$ profiles are IC versus when the $w$ profiles are a mix of IC and OOC.
    In the first case, $\bm{\Gamma}$
    has a leading eigenvector $\frac{1}{\sqrt{w}}\bm{1}$. 
    In the latter case,
    $\bm{\Gamma}$ has %$2 \times 2$
    a four block structure (ignoring the diagonal) with its leading eigenvector being different from  $\frac{1}{\sqrt{w}}\bm{1}$.
    \item 
    According to recent work on  Eigenvector Perturbation (EP), the distance between the observed  normalized leading eigenvector $\bm{v}_1$ of a sample correlation matrix $\bm{R}$ and the normalized leading eigenvector $\tilde{\bm{v}}_1$ of the population correlation matrix $\bm{\Gamma}$  is very small in $\ell_2$ norm \citep*{Morales-Jimenez}.
\end{itemize}
As a consequence of the above two, if observed $\bm{y}^{t-w+1},\cdots,\bm{y}^t$ used to compute $\bm{R}$ are all IC profiles, then $\|\bm{v}_1 - \frac{1}{\sqrt{w}} \bm{1}\|_2$ should be small.
Whereas, when $t+w-1<\tau<t$ for $\tau$ being the last monitoring 
time when the process in IC, then 
$\|\bm{v}_1 - \frac{1}{\sqrt{w}}\bm{1}\|_2$ should be large 
because some (but, not all) of these $w$ profiles used 
for computing $\bm{R}$ are  OOC. 
If $\|\bm{v}_1 - \frac{1}{\sqrt{w}}\bm{1}\|_2$ exceeds some control limit $U$, we claim at least one of the $w$ observed profiles is OOC.
Hence, our proposed control chart using this EP property needs very few assumptions about the distributions of the error/noise and the predictors $\bm{x}^k_i$. 
To determine a control limit for our control chart from an empirically useful approximation of the distribution of eigenvector perturbation under $w$ IC profiles, we use a bootstrap approach using a sample of $m$ historical profiles that are known to be generated from the IC process. 

Our simulation studies demonstrate that our control chart achieves an $ARL_0$ larger than $10^6$ under a very general class of models for the IC profiles.
As expected, such a large $ARL_0$ results in a low FAR even when the change-point occurs long after monitoring has begun (e.g., after $10^4$ observed IC profiles). 
Typically, a control chart with such a large $ARL_0$ is expected to suffer from a large detection delay.
For the competitors of our control chart, we show that a small increase of $ARL_0$ from 200 to 370 results in noticeable increases in $ARL_1$.
As a consequence, the calibrating control limits of the control charts of our competitors to achieve $ARL_0 > 10^6$  result in undesirably large $ARL_1$ values. 
Surprisingly, our proposed control chart is able to immediately detect an OOC process (i.e., $ARL_1 = 1$) even while maintaining  $ARL_0 > 10^6$. 
Moreover, the proposed control chart is computationally fast, especially when compared to its existing competition (using multiple predictors). 
The combination of the extremely large $ARL_0$, small $ARL_1$, and desirable computational speed allows this control chart to be applied to streaming data with good performance.

%Outline for paper
The paper is organized as follows. 
Section \ref{sec:background} provides a brief background on profile monitoring (especially in the nonlinear nonparametric case) and on EP theory. 
Section \ref{sec:method} constructs the EP control chart and provides our method for determining control limits. 
Section \ref{sec:simulations} covers the simulation study exhibiting the superiority of the EP control chart in specificity and sensitivity as compared to other nonparametric monitoring methods for nonlinear profiles. 
Section \ref{sec:simulations} also discusses situations in which the EP control chart may not perform satisfactorily. 
Section \ref{sec:realworld} applies the EP control chart and a PCA based control chart on a real world dataset. 
We conclude with a discussion in Section \ref{sec:futurework}.

%-----------
%-----------
\section{Background}\label{sec:background}
Here, we describe the relevant background and associated models of profile monitoring in section \ref{ssec:ProfMonBackground}, and provide a brief introduction to EP in section \ref{ssec:EigvPertBackground}.
%-----------
\subsection{Profile Monitoring}\label{ssec:ProfMonBackground}
In this paper, the model of the observed $n$ scalar quality characteristics $(y^t_1,\cdots,y^t_n)$
at each time-point $t$  is 
\begin{equation}\label{eq: intro profile} 
y^t_i = f^t(\bm{x}^t_i) + \epsilon^t_i \quad \text{for}\ i=1,\cdots,n,
\end{equation}
where  
$\bm{x}^t_i \in \mathbb{R}^d$ is the random vector of $d$ dimensional explanatory variables,  
$\bm{\epsilon}^t=(\epsilon_1^t,\cdots,\epsilon_n^t)$ 
with $\E[\bm{\epsilon}^t]=\bm{0}$ and $\Var[\bm{\epsilon}^t]=\sigma^2\bm{I}_n$ is a vector of independent and identically distributed mean-zero random noises inherently associated with the observation process. Here $\bm{I}_n$ denotes a $(n\times n)$ identity matrix.
In this setting,  we assume the random design matrices $\bm{X}^t=(\bm{x}_1^t,\cdots,\bm{x}_n^t)^\top$ at discrete monitoring time-points $t$ to be identically distributed, and   independent of the vector of errors $\bm{\epsilon}^t=(\epsilon_1^t,\cdots,\epsilon_n^t)^\top$. 
However, we allow the correlation between noises 
$\epsilon_i^r$ and $\epsilon_i^s$ at two time-points $r\neq s$ via $\Cov[\epsilon_i^r,\epsilon_i^s]=\phi \sigma^2$ with $\phi \in (-1,1)$.
We require $\bm{x}^r_i$ and $\bm{x}^s_i$ to be dependent for $r\neq s$.
For the remainder of the paper, by the abuse of notation, we denote $f(\bm{x}_i^t)$ to be the functional relationship for the IC profile, that is, $f^t=f$ if at time $t$ the process is known to be IC.

Profile monitoring typically consists of two steps: Phase I and Phase II. 
The Phase I analysis aims to define a notion of 
expected set of observations under IC process  
using a retrospective analysis of available historical IC profiles 
$\{(\bm{y}^t, \bm{X}^t):\ t=-(m-1),\cdots,0\}$ when the process is known to be in control with   $f^0=\cdots=f^{-m+1}=f$.
Using the relevant range of summary statistics for an IC process obtained from Phase I, the Phase II analysis aims to detect the unusual deviation from an IC process in an online fashion using observations 
$\{(y_i^j,\bm{x}_i^j):i\in [n]; j=-m+1,\cdots,0,\cdots,t\}$ of quality characteristics at each discrete time $t\geq 1$.
Ideally, a good profile monitoring method should not wrongly claim any process to be OOC at time $t\ge 1$ if the underlying process is actually IC at that time $t$, that is when $t\leq \tau$, where $\tau$ (typically unknown in practice) is the 
last monitoring time when the process is IC. 
If the process is IC indefinitely ($\tau = \infty)$, we measure how often false alarms occur with the mean/expected time until a false alarm, $ARL_0=\E[T_0]$, also called the IC average run length, where $T_0$ is the random time-point when the process is erroneously declared as OOC even when the process is still IC.
If $\tau<\infty$, we use the False Alarm Rate (FAR) which is the probability of the next alarm occurring being a false alarm where the monitoring scheme is reset after each false alarm (but $\tau$ remains fixed). 
Also, a good monitoring method should signal a process to be OOC at time $t>\tau$ soon after it becomes OOC at time $\tau+1$. 
Thus, another criterion of the performance of a profile monitoring method is  the 
expected/mean time-interval 
$ARL_1=\E[T_1-\tau]$ where $T_1>\tau$ is the random discrete time-point when an OOC is detected.
Typically, a control chart used for monitoring determines a process at time $t$ to be IC (or OOC) if the monitoring statistic computed using observed $\{(\bm{y}^j,\bm{X}^j)\}_{j=-m+1}^t$  at time $t$ lies inside (or outside) of an interval of lower and upper control limits.  
Akin to traditional hypothesis testing where a rejection region is specified prior to the statistical inference to ensure a desired significance level of the testing, the control limits are calibrated to ensure a desired $ARL_0$ value prior to commencing any monitoring performed in Phase II.

For existing profile monitoring methods, 
the choice of monitoring statistics  is usually based on a set of assumptions about  the underlying stochastic model of the profile and the corresponding method for estimation of $f^t$ in \eqref{eq: intro profile}. 
Performance of the parametric approaches for estimating $f^t$ depends on the assumptions about the parametric form of $f^t$ and the error distribution 
of $\epsilon_i^t$.
Poor performance of such profile monitoring methods (including incorrect $ARL_0$) can occur if the parametric assumptions regarding any of these two modeling components turn out to be incorrect.
Additionally, in practice, we may only have access to quality historical data at $t=-m+1,\cdots,0$ under IC process, and may not have any information 
about the parametric models of the OOC profiles and of the functions $f^t$ when $t>\tau$. 
To address these practical challenges, we present below a nonparametric profile monitoring method with a control chart that neither uses any parametric form of the possibly non-linear and unknown $f^t$ nor 
assumes any parametric distribution of $\epsilon_i^t$ in \eqref{eq: intro profile}.

There is extensive literature on nonparametric profile monitoring using a univariate (scalar) predictor
\citep*{
Zou2008NonparRegressionProfMon,
Colosimo2008,
qiu2010NonparProfMonByMixedEffectsModeling,
McGinnity2015, 
Qiu2018NonparDynamicCurveMon}.
The existing literature on multivariate/vector predictors is less extensive, and they can be categorized by the estimation procedures for $f^t$, the choice of monitoring statistics, the calibration methods for control limits, and whether the predictors vary over 
time. 

The methods of  \citet*{Hung2012} and \citet*{Li2019}  use support-vector regression (SVR), but the former uses an estimated $\hat{f}^t$ as the monitoring statistics along with a moving block bootstrap confidence region of ${f}^t$ as the control limit, and the later uses the nonparametric monitoring statistics of \citet{Williams2007} based on the residuals along with 
a nonparametric EWMA control chart of \citet{Hackl1991}. 
Both approaches rely on the observed ranks of the monitoring statistics compared to a set of values based on historical data. 
Although robust to possible outliers, these methods may not have good power and low $ARL_1$ because their rank-based monitoring statistics do not effectively use the underlying model 
of \eqref{eq: intro profile}. 

As an alternative, the semiparametric method by \citet*{Iguchi2021} uses an $\ell_2$-based statistic of the estimated index parameters of the Single-Index Model (SIM)  \citep{Kuchibhotla2020} for $(y_i^t,\bm{x}_i^t)$. 
Although this SIM based approach ensures a competitive FAR and $ARL_1$, estimating index parameters is computationally slow which poses a problem in control limit calibration via Monte Carlo.

Due to the sequential nature of the profile monitoring problem, calibrating to a desired $ARL_0$ needs a Monte Carlo evaluation of $ARL_0$ using simulations of random run length $T_0$, where $T_0$ is the random time-point  of encountering a false alarm. 
For example, if $T_0$  follows a geometric distribution with mean $ARL_0$, then $\Var[T_0] = O(ARL_0^2)$.
Therefore, calibrating control limits to large $ARL_0$ (say $ARL_0 > 10^6$) through a Monte Carlo approach is infeasible. 
Such approaches (as taken by \cite{Iguchi2021} and \cite{Li2019}) are not only computationally intensive but also require explicit knowledge of $f$ making the calibration effort parametric 
even when the monitoring approach for Phase II is nonparametric.

None of these existing nonparametric and semiparametric methods simultaneously enjoy a large $ARL_0$, a small $ARL_1$, and fast computational speed. 
Our proposed Eigenvector Perturbation control chart enjoys all of these desirable properties for large classes of IC and OOC profiles and with minimal assumptions about the error distributions.
Our monitoring scheme detailed in Section \ref{ssec:evecCC}
avoids the Monte Carlo approach to obtain the control limit that also achieves a large $ARL_0$ via using a Bootstrap method.
Our proposed method to calibrate control limits for the EP control chart is detailed in Section \ref{ssec:eigvcontrollimits}.

%-----------
\subsection{Eigenvector Perturbation}\label{ssec:EigvPertBackground}
Monitoring eigenspaces is popular in the change-point detection and SPC literature.
Principal component analysis (PCA) type methods have been explored by \citet*{ColosimoPacellaRoundnessProfilesfPCA2007, Viveros-Aguilera2014, WangMei2018} and others. 
To the best of our knowledge, application of $\ell_2$ distance based EP is new to the SPC literature. 
To define this EP more explicitly in a general context, let us express a square matrix
$\bm{M}\in \mathbb{R}^{w \times w}$ as $\bm{M} = \widetilde{\bm{M}} + \bm{E}$, where
$\widetilde{\bm{M}}\in \mathbb{R}^{w \times w}$ is fixed matrix and  $\bm{E}$ is some random perturbation matrix. 
Let $\bm{v}$ and $\tilde{\bm{v}}$ respectively be the leading eigenvectors of $\bm{M}$ and $\widetilde{\bm{M}}$ with corresponding eigenvalues $\lambda$ and 
 $\tilde{\lambda}$.
The goal is to answer the question: 
{\it ``Under certain assumptions about $\widetilde{\bm{M}}$ and $\bm{E}$, how `far' can the $j$th leading eigenvector ${\bm{v}}_j$ of $\bm{M}$ be from the $j$th leading eigenvector $\tilde{\bm{v}}_j$ of $\widetilde{\bm{M}}$?''}
To employ EP to profile monitoring, we must do two things: (1)
choose $\bm{M}$ appropriately to reflect IC conditions, and (2) set a control limit based on either the possible values of $\bm{E}$ or a sample of EPs under IC conditions.

For the first task, we draw inspiration from the work of \citet{Morales-Jimenez}  on  asymptotics of the eigenvalues and eigenvectors of the sample correlation matrix  $\bm{M}$  under a special case of spiked covariance model.
Although the profiles in this paper do not follow the specific model in \cite{Morales-Jimenez}, the covariance matrix of the responses from $w$ IC profiles in certain cases follow a spiked covariance model. For example, if $\bm{X}^r = \bm{X}^s$ for $r\neq s$,  then $\Cov[y^1,\dots,y^w] = \Var[f(\bm{x})] \bm{J}_w + \sigma^2 \bm{I}$ where $\Var[f(\bm{x})]$ corresponds to the common distribution of the explanatory variables $\bm{x}^t$ and $\bm{J}_w$ is a $w\times w$ unit matrix. 
If $\Var[f(\bm{x})]>\sigma^2$, then one dominant eigenvalue exists and all other eigenvalues are small and equal.
Due to the similarity of the eigenspace of the correlation matrices in our context and in the context in \cite{Morales-Jimenez}, we consider the $\ell_2$ EP of the sample correlation matrix. 
We construct the EP  control chart in Section \ref{ssec:evecCC}.  

We now discuss the background relevant to our second task of adopting EP for profile monitoring. 
For the needed control limit of the EP, the literature provides nonasymptotic bounds on how large this can be depending on the properties of $\widetilde{\bm{M}}$ and $\bm{E}$. 
An important example is the Davis-Kahan $\sin(\theta)$ Theorem \citep{Davis1970} which gives an upper bound of $\sin(\theta)$ where $\theta$ is the angle between $\tilde{\bm{v}}_j$ and $\bm{v}_j$. 
Using some arguments from trigonometry, 
any bound of $\sin(\theta)$ can be turned into a bound on $\ell_2$ perturbation $\|\tilde{\bm{v}}_j - \bm{v}_j\|_2$. 
In the past decade, there has been growing literature on improving these bounds on $\ell_2$ and, more recently, $\ell_\infty$ distances (see  \cite*{Chen2021SpectralMeth4DS} for a review).

If the aim is to only reduce the probability of false alarms, setting a control limit to one of these bounds can achieve zero probability 
of false alarms. 
Unfortunately, this and other bounds from the EP literature are not tight enough to be useful as 
practical control limits as the resulting 
control limits will have either very low or no sensitivity.
To construct our control limits in Section \ref{ssec:evecCC}, we leverage the knowledge of the existence of a nonasymptotic bound for the EP and 
the usually small perturbation of a sample correlation matrix. 

%-----------
%-----------
\section{Methodology}\label{sec:method}
In this section we describe the proposed control chart.
Section \ref{ssec:Changepointframework} details the specific profile monitoring problem we consider under a change-point framework. 
Section \ref{ssec:evecCC} details the construction of the proposed control chart, and section \ref{ssec:eigvcontrollimits} describes the procedure for obtaining a control limit based solely on a set of known, IC profiles. 
%-----------
\subsection{Change-point framework}\label{ssec:Changepointframework}
Our underlying model, 
\begin{equation}
\label{process monitor eq}
y^t_i = \begin{cases}
f(\bm{x}^t_i) + \epsilon^t_i &\text{for}\ t\leq \tau \\
h(\bm{x}^t_i)  + \epsilon^t_i & \text{for}\ t>\tau\ ,
\end{cases}
\end{equation}
of profile $\{(\bm{x}^t_i,y^t_i) : i \in [n]\}$ at time $t$ is a special case of the model in \eqref{eq: intro profile}, where $f:\mathbb{R}^d\rightarrow \mathbb{R}$ and   $h:\mathbb{R}^d\rightarrow \mathbb{R}$ are respectively the IC and  OOC functional relationships between the response $y$ and the vector of explanatory variables $\bm{x}$,  and $\tau$ is the unknown time of change-point.
In this paper, we consider  $h$ of the form 
$h = \nu f + (1-\nu) g$ 
for $\nu \in \mathbb{R}$ where the unknown $g$ is called the OOC forcing function. 
Please recall from \eqref{eq: intro profile} that the errors $\bm{\epsilon}^t$ in equations 
\eqref{eq: intro profile}-\eqref{process monitor eq} are identically distributed with $\E[\bm{\epsilon}^t]=\bm{0}$,  $\Var[\bm{\epsilon}^t]=\sigma^2\bm{I}_n$ and same correlation $\phi = \operatorname{Corr}(\epsilon_i^{r},\epsilon_i^s)$ for $r\ne s$. Again, $\bm{I}_n$ denotes the $(n\times n)$ identity matrix.
From 
either historical data or expert knowledge, 
we have $\{(\bm{X}^t,\bm{y}^t):\ t=1-m,\cdots, 0\}$ 
that are known to be $m$ noise perturbed IC profiles.
The goal is to correctly determine whether a process 
at time $T>0$ is OOC using $T$ new observed profiles $\{(\bm{X}^t,\bm{y}^t):\ t=1,\cdots, T\}$ as well as 
$m$ known IC profiles from historical data. 
That is, we wish to test the following hypothesis: 
\begin{equation}
H_0:\ \tau\ge T 
\ \text{versus}
\ H_a: \ \tau<T\ ,
\end{equation}
where $H_0$ is equivalent to $f = f^1 = \cdots = f^T$  in \eqref{eq: intro profile}.
An ideal control chart method should be able to detect a profile  being OOC whenever $T$ is close to, but greater than, $\tau$.

%-----------
\subsection{Constructing the eigenvector perturbation control chart}\label{ssec:evecCC}

% DEFINE THE MATRIX PERTURBATION SET UP
Consider the responses from $w$ observed profiles $\{\bm{y}^t\}_{t = t_0+1}^{w + t_0}$ for some $t_0$, and define $\bm{R}\in \mathbb{R}^{w\times w}$ to be the sample correlation matrix of these responses with leading eigenvector $\bm{v}$. 
 
\begin{figure}[t]
\centering
\subfloat[$\E \bm{R}$ with all profiles \quad IC][All profiles IC]{
\includegraphics[width=0.2\textwidth]{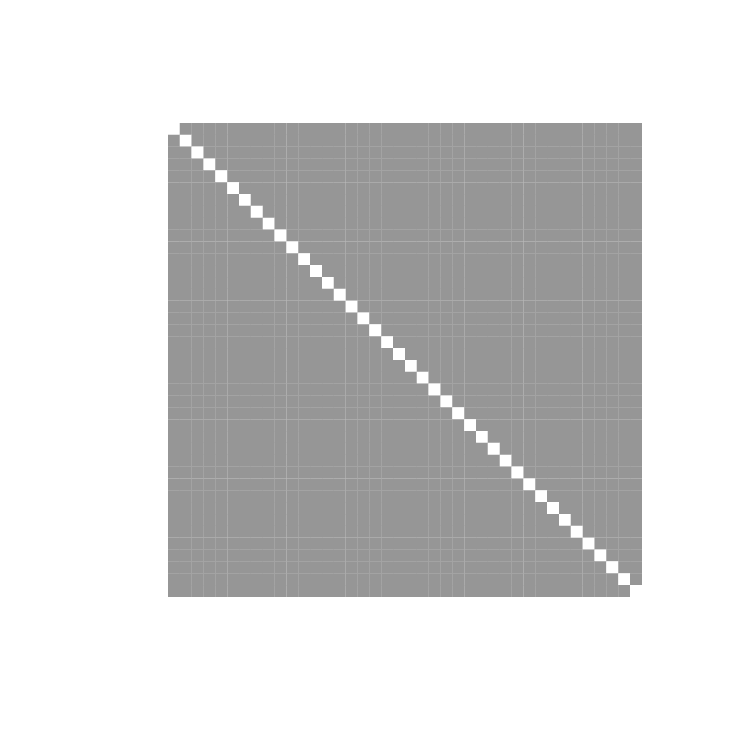}
\label{fig:visER_allIC}
}
\quad
\subfloat[$\bm{\Gamma}$ with 20 IC and 20 OOC profiles][IC and OOC profiles]{
\includegraphics[width=0.2\textwidth]{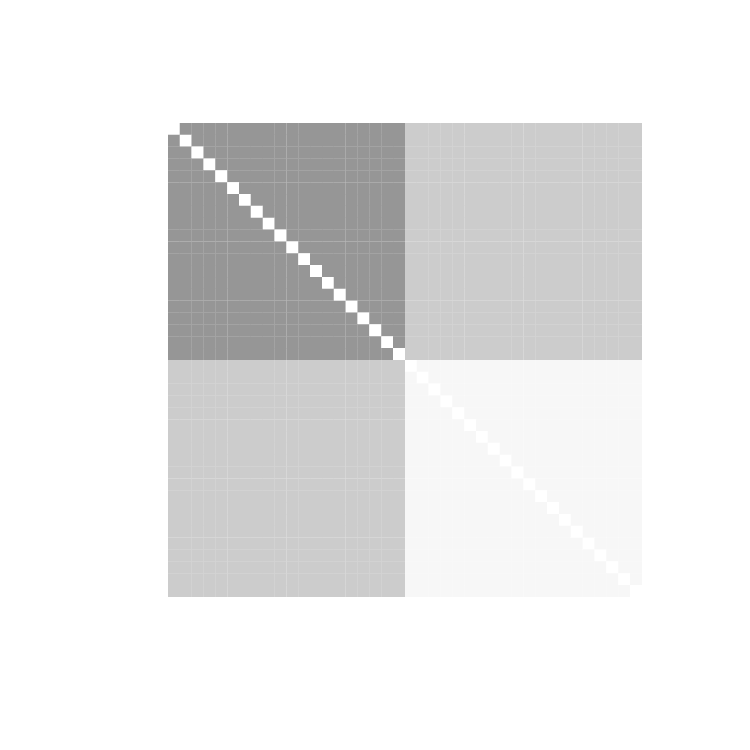}
\label{fig:visER_mixICOOC}
}
\quad
\subfloat[$\bm{\Gamma}(10)$ with 20 IC and 20 OOC profiles][$\bm{\Gamma}(10)$]{
\includegraphics[width=0.2\textwidth]{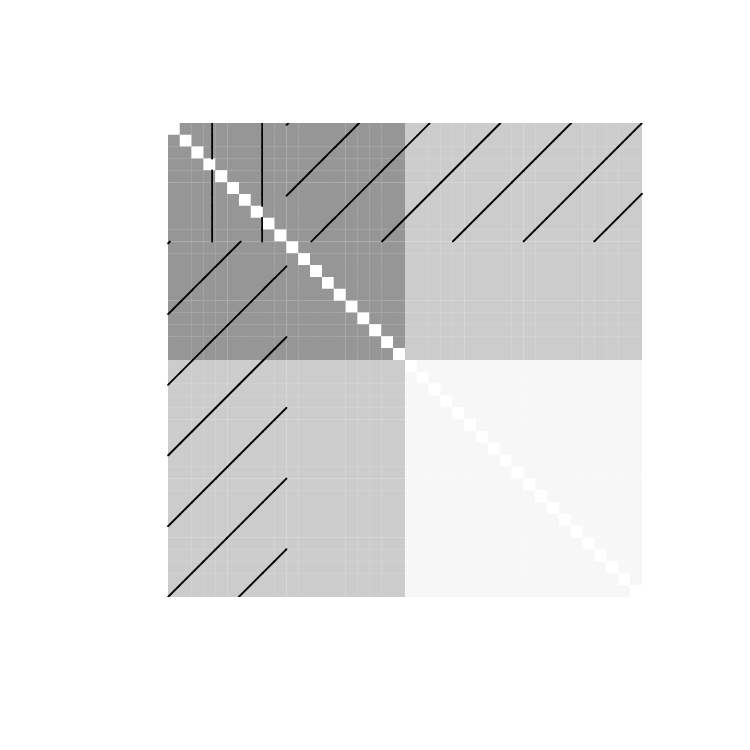}
\label{fig:visER10}
}
\quad
\subfloat[$\bm{\Gamma}(30)$ with 20 IC and 20 OOC profiles][$\bm{\Gamma}(30)$]{
\includegraphics[width=0.2\textwidth]{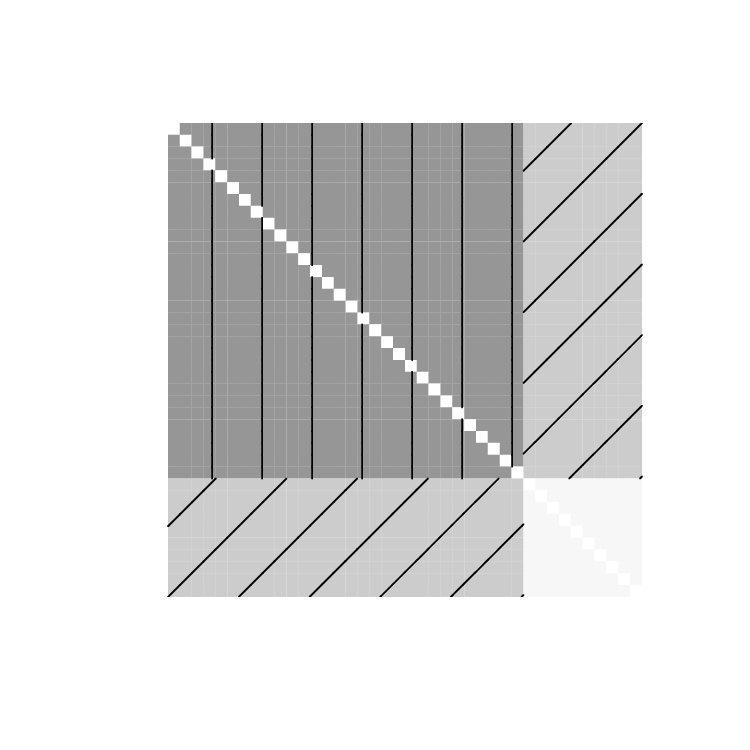}
\label{fig:visER30}
}
\caption[A visualization of correlation matrices]{
A visualization of $\bm{\Gamma}$, $\bm{\Gamma}(k_1)$ and $\bm{R}(k_1)$.
The matrix $\bm{\Gamma}$ is shown in two scenarios in (a)-(b). 
In (b)-(d), $w = 40$, $\tau = 20$, $T = 40$. 
Vertically lined regions reflect the entries substituted from the correlation matrix obtained from the historical IC profiles. 
To compute $\bm{R}(k_1)$, the entries corresponding to the diagonally lined regions need to be computed due to substituting $k_1$ of the oldest observed profiles with historical IC profiles. 
For example, 
the columns of $\bm{\Gamma}(10)$ from left to right correspond to 10 historical profiles, 10 nonhistorical IC profiles, and 20 OOC profiles. 
The correlations of the last 30 profiles with the first 10 profiles under consideration are not in $\bm{R}$ and need to be computed for $\bm{R}(10)$.}
\label{fig:visualizeER}
\end{figure}

We can consider $\bm{\Gamma}$ to be the unperturbed matrix with leading eigenvector $\tilde{\bm{v}}$ with our perturbation matrix $\bm{E}$ such that $\bm{R} = \bm{\Gamma} + \bm{E}$.
If all $w$ profiles are IC, then $\bm{\Gamma}=\bm{I}_w+(\gamma-1)\bm{J}_w$, where 
$\bm{J}_w$ is a $(w\times w)$ unit matrix, and $\gamma=\Corr[y_i^r,y_i^s]=
(\Cov[f(\bm{x}_i^r),f(\bm{x}_i^s)]+\phi\sigma^2)/(\Var[f(\bm{x}_i^r)]+\sigma^2)$ for $r\ne s$, where the $\Cov$ of the right hand side is taken with respect to the joint distribution of $(\bm{x}_i^r,\bm{x}_i^s)$.
This ``compound symmetry'' form 
of $\bm{\Gamma}$ implies that its leading eigenvector is
 $\tilde{\bm{v}} = \frac{1}{\sqrt{w}} \bm{1}$. 
However, when 
$\bm{R}$ is obtained from $k_1^*$ IC profiles and $k_2^* = w- k_1^*$ OOC profiles,
then $\bm{\Gamma}$ exhibits a block diagonal structure  as shown pictorially in Figure \ref{fig:visER_mixICOOC} 
because $\Gamma_{r,s}$ is  
\begin{equation}
\label{corrmixed}
\Gamma_{r,s}=   \Corr[y_i^u,y_i^v] = 
\frac{\Cov\left[f(\bm{x}_i^u),h(\bm{x}_i^v)\right]+
\phi\sigma^2}{\sqrt{\left(\Var[f(\bm{x}_i^u)]+\sigma^2\right) 
\left(\Var[h(\bm{x}_i^v)]+\sigma^2\right)}},
\ \text{if} 
\ r\le k_1^*<s   
\end{equation}
as $(\bm{x}_i^u,y_i^u)$ at $u$ is IC and 
$(\bm{x}_i^v,y_i^v)$ at $v$ is OOC.
Similarly, if $r,s \le k_1^*$, that is, both are IC profiles with $u,v\le\tau$, then $\Corr[y_i^u,y_i^v] = \left(\Cov[f(\bm{x}_i^u),f(\bm{x}_i^v)] +\phi\sigma^2\right)/\left(\Var[f(\bm{x}_i^u)]+\sigma^2\right)$
and $\Corr[y_i^u,y_i^v] = (\Cov[h(\bm{x}_i^u),h(\bm{x}_i^v)] +\phi\sigma^2)/\left(\Var[h(\bm{x}_i^u)]+\sigma^2\right)$ if both are OOC profiles with $u,v>\tau$.
As the row sums of $\bm{\Gamma}$ for the first $k_1^*$ rows are the same (the same is true for the last $k_2^*$ rows), $\tilde{\bm{v}} \propto \xi^* \bm{u}_1^* + \bm{u}_2^*$ where $\xi^*$ is a constant, $\bm{u}_1^* = \sum_{l = 1}^{k_1^*} \bm{e}_l$, $\bm{u}_2^* = \sum_{l = k_1^* + 1}^w \bm{e}_l$, and $\bm{e}_l$ is the $l$th standard basis vector. 

% PLOT BELOW USES SUBFIG package

% DEFINE THE MONITORING STATISTIC.
Now we leverage the differences between the leading eigenvectors under the IC scenario and under the scenario with a mix of IC and OOC profiles to create our monitoring statistic. 
Unlike under the IC condition  when $\|\bm{v} - \frac{1}{\sqrt{w}} \bm{1}\|_2$ should be small,
$\bm{v}$ should be very different from $\frac{1}{\sqrt{w}}\bm{1}$ when $w$ profiles include at least one OOC profile because in this setting $\bm{v}$ is close to $(\xi^* \bm{u}_1^* + \bm{u}_2^*)/\|\xi^* \bm{u}_1^* + \bm{u}_2^*\|_2$. 
As such, in this setting, we expect our chosen monitoring statistic $\|\bm{v} - \frac{1}{\sqrt{w}} \bm{1}\|_2$ to be large. 
Practical considerations include:
(1) an $O(wn)$ time update scheme for $\bm{R}$ between timesteps (as opposed to a na\"ive $O(w^2n)$ execution);
(2) specifying how to compute the leading eigenvector of $\bm{R}$;
(3) accounting for the possible failure of the control chart to correctly flag a profile as being OOC by time $T>\tau$.
As $\bm{R}$ is computed on a moving window, the update scheme merely computes one new row/column's worth of entries and ejects the row/column corresponding to the oldest profile in the window.
The second consideration is addressed in the supplementary materials.

Regarding the third issue, with the current procedure, all $w$ profiles considered would be OOC at time $\tau + w$, and the control chart no longer would be able to detect an OOC profile as this scenario looks identical to the one where all $w$ profiles are IC.
To mitigate this issue, we can sample without replacement $k_1$ of the $m$ historical, IC profiles to replace the $k_1> 0$ oldest profiles of the $w$ profiles being considered. 
Doing so guarantees an IC reference frame with which to compare an OOC profile.
Assuming $\bm{R}$ has already been computed for a set of $w$ profiles at time $T$, replacing $k_1$ of the oldest observed profiles modifies the sample correlation matrix. 
Denote $\bm{R}(k_1)$ be such a correlation matrix derived from $\bm{R}$ through this replacement procedure, $\bm{\Gamma}(k_1)$ be the corresponding population correlation matrix, and let $\bm{v}(k_1)$ and $\tilde{\bm{v}}(k_1)$ be the leading eigenvectors of $\bm{R}(k_1)$ and $\bm{\Gamma}(k_1)$, respectively. 
Figure \ref{fig:visualizeER} provides a visualization for two such instances of $\bm{\Gamma}(k_1)$. 
As the notation suggests, $\tilde{\bm{v}}(k_1)$ is dependent on $k_1$.
Specifically, $\tilde{\bm{v}}(k_1) \propto \xi(k_1) \bm{u}_1 + \bm{u_}2$ where $\bm{u}_1$ and $\bm{u}_2$ are defined similarly to $\bm{u}_1^*$ and $\bm{u}_2^*$.
The technical details for this derivation are in the supplementary materials.
As $\tilde{\bm{v}}(k_1)$ is a function of $k_1$, the distribution of $\bm{v}(k_1)$ is also a function of $k_1$, and hence, the distribution of $\|\bm{v}(k_1) - \frac{1}{\sqrt{w}} \bm{1}\|_2$ is also a function of $k_1$.
So a choice of $k_1$ is required, but as shown in the supplementary materials, the distribution of $\|\bm{v}(k_1) - \frac{1}{\sqrt{w}} \bm{1}\|_2$ is also a function of the two values of $\Corr[y_i^u, y_i^v]$ where $v>\tau$, which cannot be known before monitoring begins.
Therefore, $k_1$ cannot be chosen to maximize the difference between $\frac{1}{\sqrt{w}} \bm{1}$ and a normalized $\xi(k_1) \bm{u}_1 + \bm{u_}2$.
We choose instead to select $L$, evenly spaced values of $k_1$ (say $K = \left\{ 1 , \left\lfloor \frac{w}{L} \right\rfloor, 2\left\lfloor \frac{w}{L} \right\rfloor , \dots, (L-2)\left\lfloor \frac{w}{L} \right\rfloor , w-1\right\}$), and use $\max_{k_1\in K} \|\bm{v}(k_1) - \frac{1}{\sqrt{w}} \bm{1}\|_2$ as the monitoring statistic.
We combine all of the aforementioned modifications to the original monitoring scheme as detailed in the Algorithm \ref{alg:eigvCC}.

%-----------
\subsection{Determining control limits}\label{ssec:eigvcontrollimits}

The proposed control chart needs an upper control limit $U$ wherein if the monitoring statistic exceeds $U$, we claim at least one of the $w$ observed profiles to be OOC.
A nonasymptotic bound on EP found in the literature could be a candidate control limit.
Assuming we knew the quantities used in these bounds (e.g., $\|\bm{E}\|$) under IC conditions, the monitoring statistic would be guaranteed never to exceed the bound under IC conditions.
Using bounds found in the literature as control limits poses a problem: their hidden constants need to be found in order to obtain a control limit. 
These hidden constants are often too large making detecting of an OOC profile difficult.
Another method is required to determine control limits, and we use the existence of these bounds to motivate our choice of control limit.

\lstset{
    basicstyle=\linespread{0.6}\footnotesize,
    columns=fullflexible,
    breaklines=false,
    mathescape=true,
    %frame=single
    frame=bt
}
\begin{figure}[t]
\centering
\begin{lstlisting}[caption={Eigenvector Perturbation Control Chart},label={alg:eigvCC}]
Given: $\;\,$- Historical data: $\bm{X}^t,\bm{y}^t$ where $\bm{y}^t\in \mathbb{R}^n,\, \bm{X}^t\in \mathbb{R}^{n \times p} ,\,  t\in \{1-m, \dots, 0\}$
        - Upper control limit: $U>0$
        - A nonempty $K\subset [w-1]$
        - Tolerance for power iteration detector: $\zeta$
        - Window size: $w\leq m$
Do: 
Compute the sample correlation matrix $\bm{R}^\star \in \mathbb{R}^{m \times m}$ of the historical profiles.
Set $\bm{R}$ to be the last $w$ columns and rows of $\bm{R}^\star$
$S = -1$; $T$ = 1
#Conduct process monitoring
$\bf{while}$ $S < U$
    Observe $\bm{y}^T\in \mathbb{R}^{n}$ ; update $\bm{R}$ to reflect correlations of $\{\bm{y}^{T-w+1}, \dots, \bm{y}^{T}\}$
    $\bf{for}$ every $k_1$ in $K$
        $\bf{if}$ $T < w - k_1$, set $\mathcal{I} = [m - w + k_1 + T]$ $\bf{else}$ set $\mathcal{I} = [m]$
        Sample $k_1$ indices $j_1,\dots, j_{k_1}$ from $\mathcal{I}$ without replacement; set $\bm{R}(k_1)=\bm{R}$
        Replace the first $k_1$ rows/columns of $\bm{R}(k_1)$ with the $j_1,\dots, j_{k_1}$ rows/columns of $\bm{R}^\star$
        $\bm{v}(k_1)= \,\,$modified_power_iteration_detector$\left(\bm{R}(k_1),\frac{1}{\sqrt{w}} \bm{1},  \zeta\right)$
    $S = \max_{k_1\in K} \|\bm{v}(k_1) - \frac{1}{\sqrt{w}} \bm{1}\|_2$
    $\bf{if}$ $S > U$, claim change point occurred $\bf{else}$ set $T = T + 1$
\end{lstlisting}
\end{figure}

Assume a large sample of EPs $p_1, \dots, p_N$ under the IC condition are available to inform our control limit. 
Based on theory from \cite{Morales-Jimenez} and empirical observations, we can reasonably assume IC EPs to lie in an interval $(0,C]\subset(0,2)$ with $C$ small.
Ideally, we want $U$ large enough to result in a low probability of false alarm (PFA), but not so large as to be unable to detect an OOC profile.
We want to choose $U\leq C+c$ for some small $c>0$. 
If $U<C$, then PFA will be positive and hopefully small. 
If $U>C$, then $ARL_0 = \infty$, and we would want to avoid a loss of sensitivity by choosing $U$ to be close to $C$ as possible. 
We adopt an empirically useful approximation of the distribution of EPs under the IC condition to achieve this second desired effect. 
Specifically, we set $U = F^{-1}(q)$ for an aggressively large choice of $q$ where $F$ is the CDF of a normal distribution fitted to the $p_1,\dots, p_N$ using typical moment estimates.
The aggressive choice of $q$ will result in a large $U$, and the light tail of a Gaussian will limit how large $U-C$ can become should the choice of $c$ result in $U>C$. 
As the maximum of a set of EPs also shares similar boundedness properties as the $N$ EPs, we apply this ``quantile trick'' to a large set of IC monitoring statistics $S_1, \dots, S_N$. 
Bootstrapping can be used to obtain the required large sample of monitoring statistics, with one such approach detailed in the supplementary materials.
%
%-----------
%-----------
\section{Simulations}\label{sec:simulations}
In this section, we detail results from simulation studies. 
The first simulation study compares the EP control chart against existing nonparametric and semiparametric profile monitoring methods.
The second simulation study aims to identify conditions that degrade performance of our method. 
Supplementary materials address technical details regarding control limit calibration, scenario details, two competitors' performances, an illustrative example using distinct, dependent predictors between time steps, and effect size calibration.  

%-----------
\subsection{The general setup for the simulation studies}\label{ssec:setup}
Each simulation study consists of a set of scenarios that typically cover all possible combinations of factors of interest.  
For each scenario (simulation model), we simulate 100 replications (trials) of the sequence of profiles from the underlying simulation model and the corresponding process monitoring to obtain the Monte Carlo estimates of the $ARL$ and FAR of each method.
For each replication/trial, $m$ historical profiles were randomly generated, and a control limit was set.
How the control limit was determined was dependent on the monitoring scheme used.
Once a control limit was established, we began monitoring, and a new profile was generated for each time step. 
If a false alarm occurred, the control chart was restarted as if monitoring had just begun. 
The time of the last IC profile $\tau$ was fixed remaining unaffected by the number of false alarms. 
Monitoring ended either when the control chart raised an alarm after time $\tau$ or if it failed to provide a true alarm after time $T_{\text{timeout}}$.
If $t^*_{i}$ denotes the time when a true alarm occurs for the $i$th trial, the Monte Carlo estimate of $ARL_1$ is $\frac{1}{100} \sum_{i=1}^{100} (t^*_{i} - \tau)$, assuming that all 100 trials terminated prior to $T_{\text{timeout}}$. The estimate of the FAR is the proportion of alarms that are false alarms. 

We need to consider the effects various types of functions $(f,h)$ have on the performance of profile monitoring methods in order to compare them. 
We use a signal-to-noise ratio (SNR) as our primary choice of effect size, though other related effect sizes are considered.
We define the SNR as $\Var[f(\bm{x})-h(\bm{x})] / \sigma^2$,  and the variance is taken with respect to a random $\bm{x}$.
For the remainder of the paper, we refer to $\Var[f(\bm{x})-h(\bm{x})]$ as $\Var[f-h]$ for brevity.
Similarly, we refer to $\Cov[f(\bm{x}), h(\bm{x})]$ as $\Cov[f,h]$.
Unless otherwise stated, we use $\bm{\epsilon}^t \sim N(\bm{0},\bm{I})$ for all $t$, so the SNR is simply $\Var[f-h]$.
Also unless otherwise stated, we assume the OOC functional relationship $h$ to be an affine combination of $f$ and some OOC forcing function $g$, where the affine combination parameter is chosen to achieve the desired effect sizes such as SNR and between-profile correlation.

%-----------
\subsection{Comparing with other approaches}\label{ssec:evecwins}
We compare the proposed EP control chart with competing nonparametric control charts in this section and demonstrate its superior performance. 
The competing control charts are the SIM based approach from \cite{Iguchi2021}, the SVR method of \cite{Li2019}, a wavelet based method of \cite{Chicken2009}, and a PCA based control chart from \cite{Colosimo2010Comparison} hereinafter referred to as the PCA control chart.
We also implemented the methods of \cite{Zou2008NonparRegressionProfMon} and  \cite{qiu2010NonparProfMonByMixedEffectsModeling}, but we place their results in the supplementary materials due to computational issues they faced preventing monitoring from occurring.
We only use the simulation models where error distributions are independent over time with $\Corr[\epsilon_i^r,\epsilon_i^s]=0$, and common $\bm{x}_i^t=\bm{x}_i$ for all time-points $t$ are sampled independently for $i=1,\cdots,n$ from a common $p$-dimensional multivariate distribution. 
So, the expression of $\Gamma_{r,s}$ in (\ref{corrmixed}) is only $\Cov[f(\bm{x}),h(\bm{x})]/\sqrt{\Var[f(\bm{x})]\Var[h(\bm{x})]}$ if one is from IC and another from OOC profiles. 
Even though our method is equipped for process monitoring under a very general set-up, especially when $\bm{x}_i^t$ are different over time and $\Corr[\epsilon_i^r,\epsilon_i^s]=\phi\ne 0$ for $r \ne s $, we use this simpler simulation model because we would like to make a fair comparison with competing methods which were not meant to accommodate these general modeling assumptions of our method.
% -----------
\subsubsection{Multivariate predictors}\label{ssec:multivariatewin}
\begin{table}[]
\centering
\caption{Performance of non/semi-parametric profile monitoring with multiple predictors. 
}
\scalebox{0.8}{
\begin{tabular}{llllllll}
Method &
  $ARL_0$ &
  $\tau$ &
  \begin{tabular}[l]{@{}l@{}}Range of observed FAR\end{tabular} &
  \begin{tabular}[l]{@{}l@{}}Range of  observed $ARL_1$\end{tabular}  &
  \begin{tabular}[c]{@{}l@{}}Fast Calibration?\end{tabular}   &
  \\ \hline
\multirow{2}{*}{\citet{Li2019}}         & 200   & 30        & $(0.17 , 0.37)$   & $(4.24, 4.86 )$   & No   \\
                                        & 370   & 30        & $(0.07,0.34)$     & $(4.37, 5.39)$    & No   \\ \hline
\citet{Iguchi2021}                      & 200   & 30        & $(0.07 , 0.25)$   & $(1,2.44 )$       & No   \\
                                        & 370   & 30        & $(0.01,0.11)$     & $(1,2.75)$        & No   \\ \hline
Eigenvector                             & 200   & 30        & $(0.17, 0.69)$    & $( 1,1 )$         & No   \\
Perturbation                            & 370   & 30        & $(0.01, 0.55)$    & $( 1,1 )$         & No   \\
                            & $>5 \times 10^6$  & 30        & $(0,0)$           & $( 1,1 )$         & Yes  \\
                            & $>5 \times 10^6$  & $10^4$    & $(0,0.01 )$       & $( 1,1 )$         & Yes 
\end{tabular}
}
\label{tab:Eigv_wins_ARL1}
\end{table}
We begin with comparing nonparametric profile monitoring methods with multiple predictors. 
The profile monitoring methods were used on each scenario of a full-factorial design of factors and levels taken from \citet{Iguchi2021}. 
We summarize the results in Table \ref{tab:Eigv_wins_ARL1} by listing the range of FAR and $ARL_1$ across all scenarios of the experiment with noninteger values being rounded. 

The PCA control chart results are not listed as it failed to raise a true alarm by $T_{\text{timeout}}=11500$ in 
3195 of the 3200 trials conducted across 32 scenarios.
Observe the proposed method using a control limit obtained from bootstrapping outperformed its competitors: 
the sensitivity of the proposed method is not affected by dramatically increasing the $ARL_0$. 
Lower bounds for the $ARL_0$ of the EP control chart using the control limit obtained through bootstrapping are listed in the supplementary materials. 
The competing methods either cannot calibrate to such a large $ARL_0$ or underperform with respect to FAR and $ARL_1$. 
If the SIM and SVR methods could be calibrated to an $ARL_0 > 5 \times 10^6$, the FAR could certainly be competitive with the EP control chart.
Doing so, however, comes at the cost of increasing $ARL_1$ as demonstrated by increasing the $ARL_0$ from 200 to 370.
Even at $ARL_0 = 370$, both of these competing methods could not achieve an $ARL_1 = 1$ across all scenarios in the computer experiment, while the EP control chart achieved this feat.
As a low FAR should be expected from setting $\tau = 30$ when a control chart is calibrated to $ARL_0 > 5 \times 10^6$, we increased $\tau$ to $10^4$ for the proposed method.
Even at a large value of $\tau$, the FAR is extremely small.
In fact, 13 of 16 scenarios resulted in no false alarms.

% -----------
\subsubsection{Univariate predictors}\label{ssec:univariatewin}
As there are competing methods designed with one predictor in mind, we compare them against the proposed method with a single predictor. 
The smallest lower bound on the observed $ARL_0$ for the EP control chart in Section \ref{ssec:multivariatewin} was 5010603 so the PCA control chart was calibrated to such an $ARL_0$ in addition to $ARL_0$ of 200 and 370.
We set $T_{\text{timeout}}$ to be 10500, 11500, and 3000 for the EP, PCA, and wavelet control charts, respectively. 
\begin{table}[h]
\centering
\scalebox{0.6}{
\begin{tabular}{cc|ccc|cc|ccc|}
\cline{3-10}
 &  & \multicolumn{3}{c|}{PCA} & \multicolumn{2}{c|}{Wavelet} & \multicolumn{3}{c|}{EP ($w \in \{20, 40\}$)} \\ \cline{2-10} 
\multicolumn{1}{c|}{} & \diagbox[width=5em]{$ARL_0$}{$\tau$} & \multicolumn{1}{c|}{0} & \multicolumn{1}{c|}{30} & $10^4$ & \multicolumn{1}{c|}{0} & 30 & \multicolumn{1}{c|}{0} & \multicolumn{1}{c|}{30} & $10^4$ \\ \hline
\multicolumn{1}{|c|}{\multirow{3}{*}{\begin{tabular}[c]{@{}c@{}}Range of\\ observed \\ FAR\end{tabular}}} & 200 & \multicolumn{1}{c|}{-} & \multicolumn{1}{c|}{(0, 0)} & (0, 0) & \multicolumn{1}{c|}{-} & $( 0.04, 0.15  ) $ & \multicolumn{1}{c|}{-} & \multicolumn{1}{c|}{*} & * \\ \cline{2-10} 
\multicolumn{1}{|c|}{} & 370 & \multicolumn{1}{c|}{-} & \multicolumn{1}{c|}{(0, 0)} & (0, 0) & \multicolumn{1}{c|}{-} & $( 0.04, 0.15  ) $ & \multicolumn{1}{c|}{-} & \multicolumn{1}{c|}{*} & * \\ \cline{2-10} 
\multicolumn{1}{|c|}{} & ``large'' & \multicolumn{1}{c|}{-} & \multicolumn{1}{c|}{(0, 0)} & (0, 0) & \multicolumn{1}{c|}{-} & * & \multicolumn{1}{c|}{-} & \multicolumn{1}{c|}{$(0 , 0) $} & $(0, 0.07 ) $ \\ \hline
\multicolumn{1}{|c|}{\multirow{3}{*}{\begin{tabular}[c]{@{}c@{}}Max \\ observed \\ $ARL_1$\end{tabular}}} & 200 & \multicolumn{1}{c|}{$\infty$} & \multicolumn{1}{c|}{$\infty$} & $\infty$ & \multicolumn{1}{c|}{3.34} & 447.27 & \multicolumn{1}{c|}{-} & \multicolumn{1}{c|}{*} & * \\ \cline{2-10} 
\multicolumn{1}{|c|}{} & 370 & \multicolumn{1}{c|}{$\infty$} & \multicolumn{1}{c|}{$\infty$} & $\infty$ & \multicolumn{1}{c|}{3.37} & 455.20 & \multicolumn{1}{c|}{-} & \multicolumn{1}{c|}{*} & * \\ \cline{2-10} 
\multicolumn{1}{|c|}{} & ``large'' & \multicolumn{1}{c|}{$\infty$} & \multicolumn{1}{c|}{$\infty$} & $\infty$ & \multicolumn{1}{c|}{-} & * & \multicolumn{1}{c|}{12.70} & \multicolumn{1}{c|}{12.97} & 12.86 \\ \hline
\multicolumn{1}{|c|}{\multirow{3}{*}{\begin{tabular}[c]{@{}c@{}}Scenarios \\ with \\ $ARL_1 =1$\end{tabular}}} & 200 & \multicolumn{1}{c|}{12/16} & \multicolumn{1}{c|}{12/16} & 12/16 & \multicolumn{1}{c|}{13/16} & 13/16 & \multicolumn{1}{c|}{-} & \multicolumn{1}{c|}{*} & * \\ \cline{2-10} 
\multicolumn{1}{|c|}{} & 370 & \multicolumn{1}{c|}{13/16} & \multicolumn{1}{c|}{13/16} & 13/16 & \multicolumn{1}{c|}{13/16} & 13/16 & \multicolumn{1}{c|}{-} & \multicolumn{1}{c|}{*} & * \\ \cline{2-10} 
\multicolumn{1}{|c|}{} & ``large'' & \multicolumn{1}{c|}{10/16} & \multicolumn{1}{c|}{10/16} & 10/16 & \multicolumn{1}{c|}{-} & * & \multicolumn{1}{c|}{20/32} & \multicolumn{1}{c|}{20/32} & 20/32 \\ \hline
\end{tabular}

}
\caption{A comparison of nonparametric profile monitoring methods with a single predictor}
\label{tbl: compare univariate}
\end{table}

We summarize the results in Table \ref{tbl: compare univariate}. 
Although the PCA control chart was optimal in $ARL_1$ in some scenarios, it often was unable to raise a true alarm. 
The wavelet control chart detected an out-of-control profile in all scenarios, but as $\tau$ increased, the $ARL_1$ greatly increased (e.g., from 3.37 to 446) in six scenarios. 
The wavelet control chart leveraging all previously observed profiles explains the $O(T^2)$ computational complexity at time $T$ (and the reason for omitting calibrating to $ARL_0 = 5010603$) and may also explain the dependence of its $ARL_1$ on $\tau$.
The EP control chart using the ``quantile trick'' should result in an extremely large $ARL_0$ per the simulation study in Section \ref{ssec:evecwins} and yields respectable FAR and $ARL_1$. 
We claim the EP control chart is preferable to the two alternatives as multiple predictors are allowed, long run lengths are feasible, and achieves good performance.
%-----------
\subsection{Investigating the boundaries of ``good'' performance}\label{ssec:toughscenarios}
Given the promising results thus far, we now investigate settings which are difficult for the EP control chart. 
In Section \ref{ssec:quadquad}, we find combinations of control chart parameters and properties of the IC and OOC profiles which are difficult for the EP control chart. 
The EP control chart loses its good performance under conditions which are the result of windowing, a consequence of monitoring correlations, or truly difficult for any nonparametric profile monitoring procedure. 
We show via simulation robustness to moderate levels of within profile correlated errors, small SNRs, and heavy tailed errors in Section \ref{ssec:othertoughscenarios}.
%----
\subsubsection{Parameters of the control chart and the profiles}\label{ssec:quadquad}
In the previous subsection, we demonstrated the superior performance of the EP control chart.
In this subsection we explore the scenarios where the performance of the EP control chart degrades and no longer achieves a zero FAR and $ARL_1$ of one.

There are some scenarios in which we should expect the EP control chart to fail to detect an OOC profile such as $h \propto f$ or $h = f + c$ for some constant $c$.
We consider the first such difficult scenario. 
As such, we consider 
the correlation $\rho[f,h]$ between the IC and OOC functional relationships, $f$ and $h$, respectively.
Additionally, the variance of the sample correlation of 
$\bm{y}^r, \bm{y}^s$, with $r,s\leq \tau$
depends on $\Var[f]/\sigma^2$, so $\Var[f]$ is also included.

\begin{figure}[ht]
    \centering
    \includegraphics[width =  0.54\linewidth]{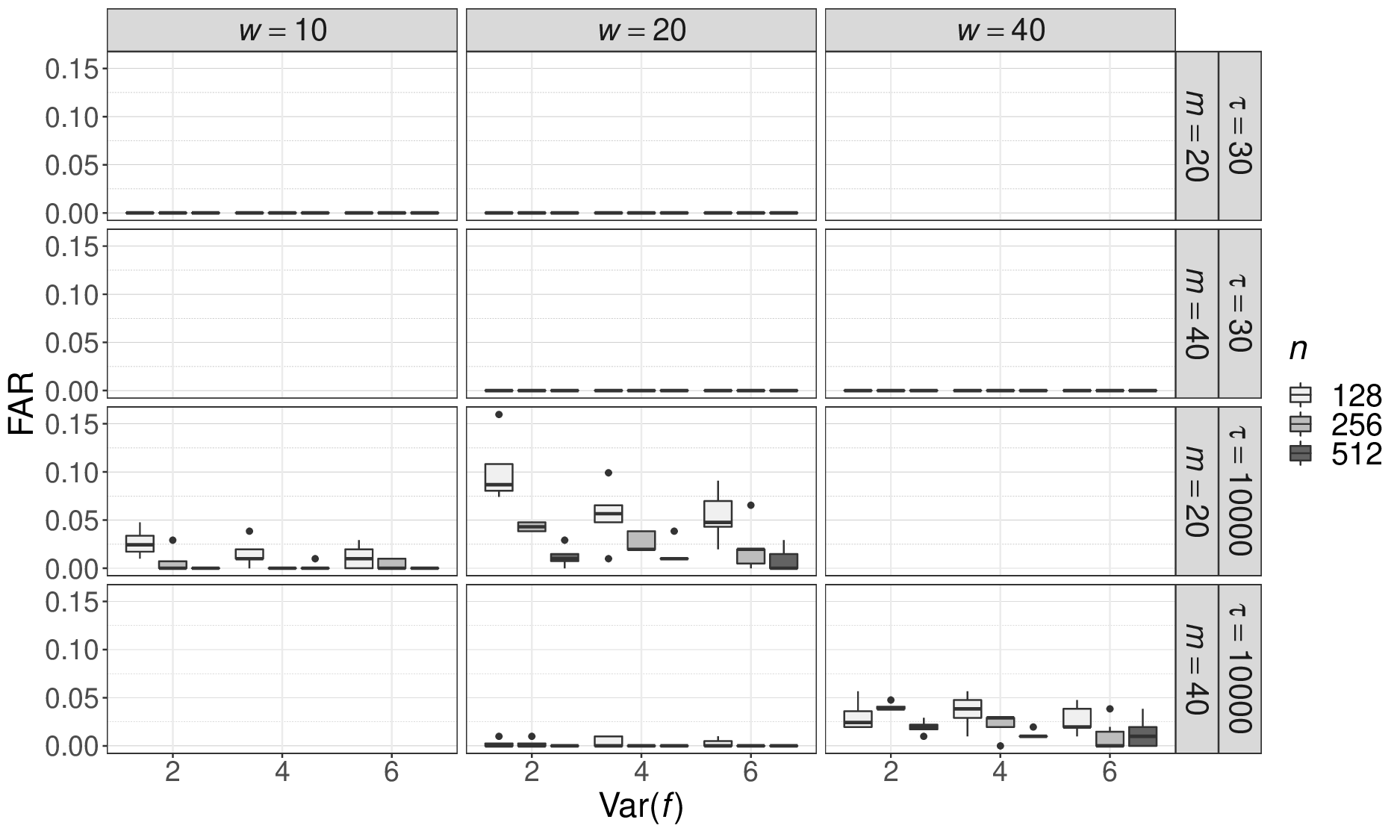}
    \includegraphics[width =  0.45\linewidth]{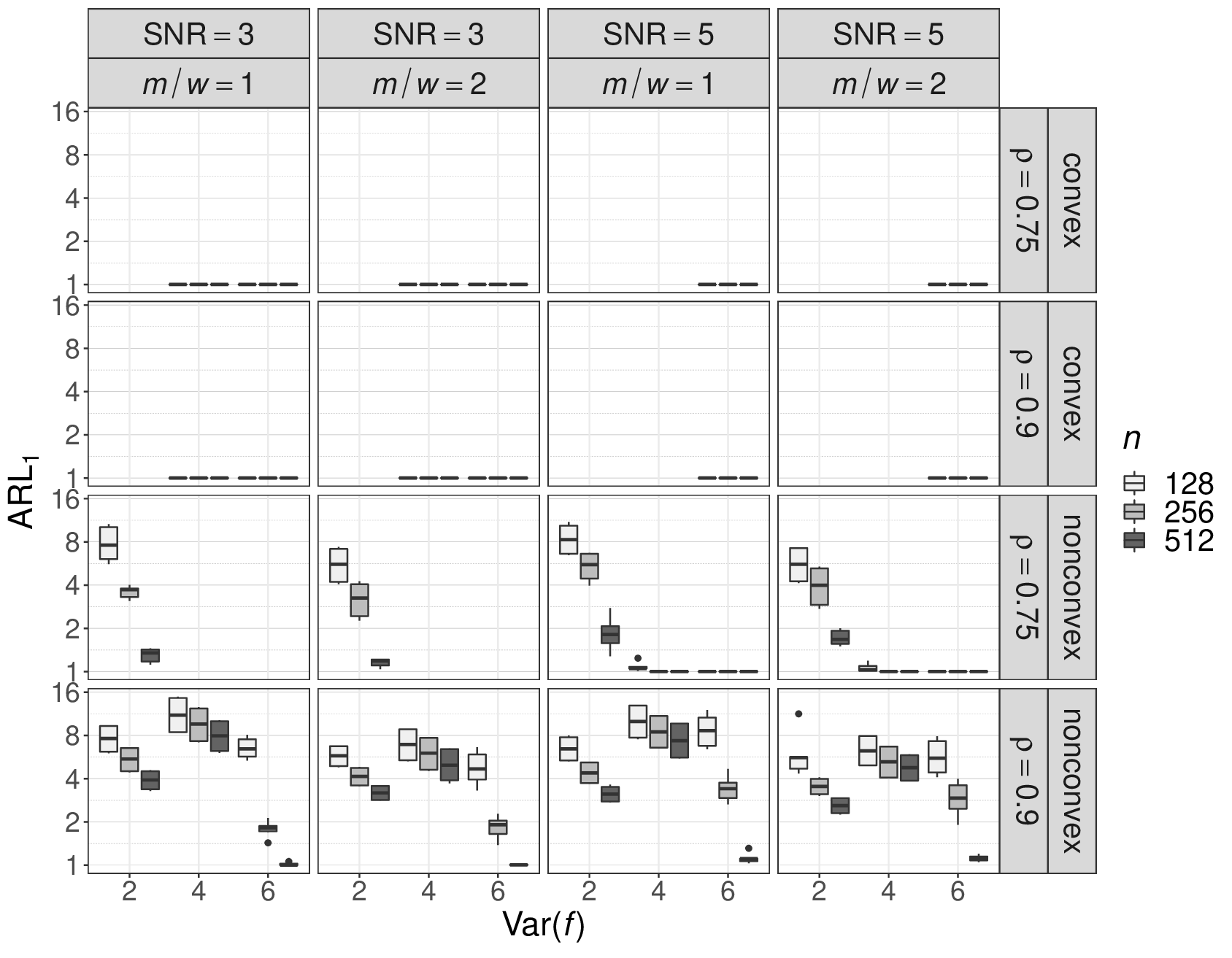}
    \caption{
    Boxplots of the FARs (left) and $ARL_1$ (right) for the proposed control chart for the simulation study in Section \ref{ssec:multivariatewin}. 
    Each boxplot on the left is drawn using the FARs across the various choices of SNR, $\rho[f,h]$, and choice of a convex combination of $f$ and $h$, as these quantities affect alarms after $\tau$ and do not affect the FAR. 
    Each boxplot on the right is drawn using the $ARL_1$ across the various choices of $\tau$ and $m$.
    These factors are omitted in the facet plot as $\tau$ and $m$ have little to no effect on $ARL_1$. 
    Flat boxplots with no outliers indicate instances of zero false alarms (left) or $ARL_1=1$ (right) across all relevant treatment combinations. 
    Scenarios not in the experimental design result in empty plots. 
    }
    \label{fig:quadquad_both}
\end{figure}

In this subsection we consider profiles where the functional relationship in a profile are zero intercept quadratic polynomials.
That is, we assume $f, g,$ and $h$ are of the following form: $
\bm{x}^\top \bm{A} \bm{x} + \bm{a}^\top \bm{x}$
where $\bm{A} \in \mathbb{R}^{d\times d}$, $\bm{a} \in \mathbb{R}^d$.
The factors and levels of the simulation study are outlined in the supplementary materials. 
For the simulation study, $\bm{X}$ was randomly chosen from $\operatorname{Unif}([0,1]^{25})$ at the beginning of each trial and held fixed for all time steps $t$. 
Scenarios consisted of all possible combinations of $\tau\in\{0,30,10^4\}$, $\log_2(n) \in \{7,8,9\}$, $m\in \{20,40\}$, $m/w\in \{1,2\}$, $\text{SNR}\in\{3,5\}$, $\Var[f]\in \{2,4,6\}$, $\rho(f,h)\in \{0.75,0.9\}$, and the convexity of $h$ with respect to $f$ and $g$ (i.e., if $\nu \in [0,1]$). 
Figure \ref{fig:quadquad_both} shows the estimates of the FAR across the relevant factors in the experimental design which affect IC performance. 
As expected, the  FAR for $\tau = 30$ was zero in all but one case.
When $\tau = 10^4$, we observe excellent performance when $n = 512$. 
Also notice as $m/w$ increased, the FAR decreased.
Finally, as $\Var[f]$ increased with $n$ fixed, the FAR tended to decrease.
Therefore, a practitioner can decrease the FAR by increasing $n$, $m$, $m/w$, and, if possible, $\Var[f]/\sigma^2$. 

Figure \ref{fig:quadquad_both} depicts the approximations of the $ARL_1$ across the relevant factors in the experiment which influence OOC performance.
As expected, $ARL_1$ decreased with increasing $n$, $m/w$, or $\rho(f,h)$.
The $ARL_1$ did not decrease uniformly with increasing SNR and $\Var[f]$, and this counterintuitive behavior is a result of some nonintuitive relationships between SNR, $\Var[f]$, and $\rho[f,h]$ which is explained in the supplementary materials. 

We should expect poor performance when $n$ is small, $\rho[f,h] \approx 1$, and $\nu>1$.  
For example, when $n=64$ $\rho[f,h] =0.9$, $\Var[f]<6$, and $\nu >1$, 30 trials of 4800 trials across the relevant 48 treatment combinations in the experimental design failed to raise an alarm by time $t = 10500$.  
Additionally, we repeated the same simulation experiment but used smaller and negative correlations (with $\rho[f,h] \in \{ -1, -0.75, \pm0.5, \pm0.3, 0\}$).
For all of these additional settings of $\rho[f,h]$, the control chart demonstrated optimal performance with $ARL_1 = 1$. 

%----
\subsubsection{Other difficult scenarios}\label{ssec:othertoughscenarios}
\begin{figure}[ht]
    \centering
    \includegraphics[width = 0.8\linewidth]{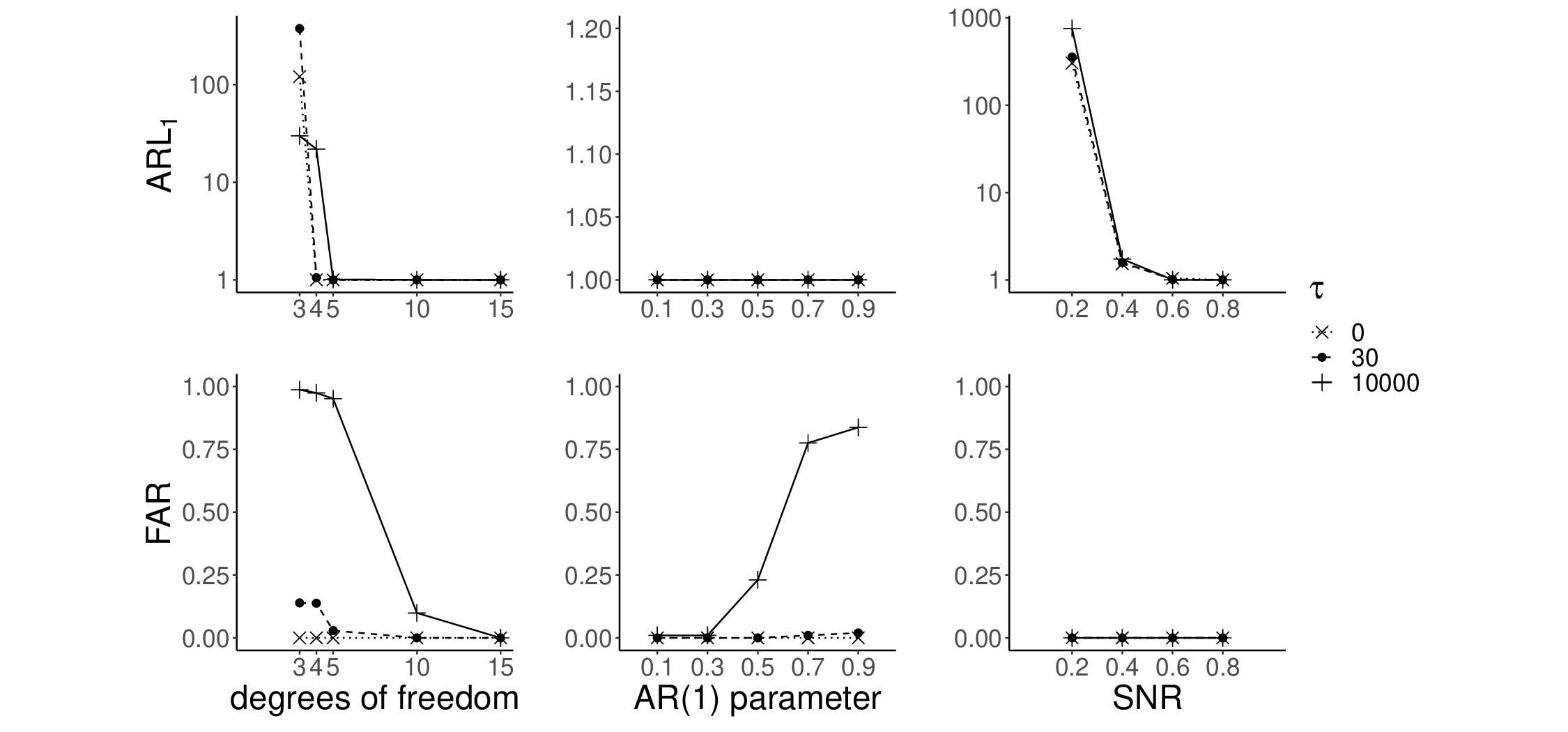}
    \caption{EP control chart performance in difficult scenarios. 
    (left) Errors are t-distributed.
    (middle) Within profile correlation with AR(1) errors.
    (right) SNR$<1$ with $\Var[f] = 6$.}
    \label{fig:othertoughcases}
\end{figure}

In this subsection, we consider three other difficult scenarios and present the results of these simulations in Figure \ref{fig:othertoughcases}. 
\newtextforblind{\blind}{The first scenario uses heavy tailed errors (i.e., t-distributed with varying degrees of freedom).}
The performance of the EP control chart is unsatisfactory when the errors have \newtextforblind{\blind}{less than five moments},
but begins to achieve good performance when the errors have \newtextforblind{\blind}{more} than four moments. 
This transition in performance agrees with the theory in EP for covariance and correlation matrices which typically assumes the data have greater than four moments. 
The second scenario involves within-profile correlation. 
We place an AR(1) structure on the errors given their order of appearance within a profile (i.e., $\Corr[\epsilon_i,\epsilon_j] = \psi^{|i-j|}$). 
We see the EP control chart can handle moderate levels of within-profile correlation as measured by the autocorrelation parameter. 
The third scenario considers a case where the SNR is small but $\Var[f]$ is large. 
The EP control chart only performs poorly once the SNR reaches a value of 0.2, which is exceptionally small. 

%-----------
%-----------
\section{Real World Dataset Application}\label{sec:realworld}

\newtextforblind{\blind}{
We apply the EP and PCA control charts on a robot arm approaching and grasping an object before trasferring it to a new location and ungrasping it
\citep{Dua2019,Camarinha-Matos1996}.}
Force and torque measurements are collected along three axes every 315 ms for 4.725s under 
\newtextforblind{\blind}{successful and failed attempts.}
\newtextforblind{\blind}{
The supplementary materials contain  visualizations and details on the dataset and its IC and OOC scenarios.}

\newtextforblind{\blind}{
The robotic dataset is well suited for demonstrating our EP control chart due to its unique combination of small sample sizes and large variance ratio, $\Var[f]/\sigma^2$. 
Using timestamps and measurement types as predictors, which do not change between attempts, an executed task results in a profile with $n=90$.
With only 18 of the 21 normally executed tasks being considered to form IC profiles after data cleaning, $m$, $n$, and $w$ are relatively small implying poor EP control chart performance unless $\Var[f]/\sigma^2$ is exceptionally large. 
The observed variance ratio under normal operating conditions meets this condition with $\widehat{\Var}[\hat{f}]/\hat{\sigma}^2 = 72.8$. 
Having such a large variance ratio is intuitive for robotic arm tasks: the variability between repeated attempts is minimal due to the precision of robotic movements, resulting in a small $\sigma^2$. 
At the same time, tasks requiring substantial movement produce large variability in sensor measurements, leading to large $\Var[f]$. 
As such, the robot arm dataset provides an ideal context where the EP control chart can excel despite small $m$, $n$, and $w$.
}

\newtextforblind{\blind}{For simulations, we used $(w,m)\in \{4,5,6\}\times\{11,12,13\}$, $\tau = 18 - m$, and
three different types of collision with the robot arm to form OOC scenarios.}
For each trial, we randomly permute the IC profiles, assign the first $m$ to be historical IC profiles, and let the rest be observed in an online fashion. 
The OOC profiles are sampled without replacement. 

\newtextforblind{\blind}{Using t}he PCA control chart no false alarms were raised, but we often observed $ARL_1>1$. 
The $ARL_1$ averaged over all of the scenarios were 1.46, 1.53, and 2.12 for $ARL_0$ of 200, 370, and $5010603$, respectively.
Using the EP control chart, $ARL_1 = 1$ was observed for all scenarios.
\newtextforblind{\blind}{Only four of the 27 scenarios simulated resulted in positive FARs, all less than 0.02, occurring when $(w, m) \in \{(4, 11), (4, 12), (5, 11), (6, 11)\}$.
As seen in Section \ref{ssec:multivariatewin}, we suspect the small $w$ and $m$ contribute to the positive FAR despite the small $\tau$.  
}
%-----------
%-----------
\section{Discussion}\label{sec:futurework}

We established a new nonparametric profile monitoring 
tool using EP to deal with a vector of quality characteristics that are a scalar function  of multiple explanatory variables. 
For existing nonparametric profile monitoring methods 
for this situation, a Monte Carlo approach is commonly required to calibrate control limits. 
For such approaches, calibrating to 
large $ARL_0$ on the order of $10^6$ is infeasible and impractical. 
Unlike the profile monitoring in the literature that typically calibrates control limits to $ARL_0$ of 200 or 370, our
proposed profile monitoring method can quickly calibrate to a very large $ARL_0 > 10^6$. 
Usually, the $ARL_1$ increases as the $ARL_0$ increases. 
Despite achieving such a large $ARL_0$, our EP based control chart suffers little loss in $ARL_1$. 
\newtextforblind{\blind}{The success of the EP control chart lies in the sensitivity of $\|\bm{v} - \frac{1}{\sqrt{w}} \bm{1}\|$ to changes from a one-block structure of $\bm{R}$ and the ability to obtain large specificity (by decreasing variability of the entries of $\bm{R}$) with sufficiently large $n$ or $\Var[f]/\sigma^2$.
Large values of $m$ and $m/w$ help in decreasing the FAR and increasing the efficiency of the bootstrapping used to find control limits.  
Performance degrades under truly difficult situations such as when $n$, $m$, $\Var[f]/\sigma^2$, or $\left|\Corr[f(\bm{x}_i^r),f(\bm{x}_i^s)] -  \Corr[f(\bm{x}_i^r),h(\bm{x}_i^t)]\right|$ for $r,s\le \tau<t$ 
are small.
The robot failures application shows increasing one of these quantities can compensate for a small value of one of the other quantities. 
}

Future work may include monitoring not just correlations but also if $\nu = 1$.
The regime of $(n,m)$ for our method contrasts with methods requiring $m$ to be large ($\geq 500$) and allowing $n$ to be small (e.g., \cite{qiu2010NonparProfMonByMixedEffectsModeling}).
Achieving similar performance for small $n$  remains an open problem.
It is not uncommon for time series data to be autocorrelated as opposed to the equicorrelated structure explored in this paper, so the proposed EP control chart could be modified to handle a different correlation structure. 
Relaxing the need for dependent predictors $\bm{x}^r, \bm{x}^s$, $r\neq s$ would also be valuable.

\section*{Competing Interests}
The authors report there are no competing interests to declare.

\bigskip
\begin{center}
{\large\bf SUPPLEMENTARY MATERIAL}
\end{center}

\begin{description}

\item[Supplementary manuscript:] Contains miscellaneous mathematical details, pseudocode for the proposed procedures, the experimental designs used in the simulation studies,  and miscellaneous simulation results.

\item[Code for the simulation study using quadratic profiles] R code to execute the EP control chart in the simulation study in Section \ref{ssec:othertoughscenarios}. 

\end{description}

\bibliographystyle{Chicago}
% \bibliography{references}  %%% Remove comment to use the external .bib file (using bibtex).

\end{document}

\def\spacingset#1{\renewcommand{\baselinestretch}%
{#1}\small\normalsize} \spacingset{1}

%%%%%%%%%%%%%%%%%%%%%%%%%%%%%%%%%%%%%%%%%%%%%%%%%%%%%%%%%%%%%%%%%%%%%%%%%%%%%%

\if0\blind
{
  \title{\bf Supplementary Material: Profile Monitoring via Eigenvector Perturbation}
  \date{November 6, 2024}
  \author{Takayuki Iguchi\thanks{
    T. Iguchi has changed affiliation to the Air Force Institute of Technology, but the work was performed while at Florida State University. 
    This work was supported by the Office of the Secretary of Defense, Directorate of Operational Test and Evaluation under Grant FA8075-14-D-0019 and the Test Resource Management Center under the Science of Test research program under Grant FA807518F1525.
    The views expressed in this article are those of the author and do not reflect the official policy or position of the United States Air Force, Department of Defense, or the U. S. Government.
    }\hspace{.2cm}\\
    Department of Statistics, Florida State University\\
    and \\
    Andr\'{e}s F. Barrientos \hspace{.2cm}\\
    Department of Statistics, Florida State University\\
    and \\
    Eric Chicken \hspace{.2cm}\\
    Department of Statistics, Florida State University\\
    and \\
    Debajyoti Sinha \hspace{.2cm}\\
    Department of Statistics, Florida State University}
  \maketitle
} \fi

\if1\blind
{
  \bigskip
  \bigskip
  \bigskip
  \begin{center}
    {\LARGE\bf Supplementary Material: Profile Monitoring via Eigenvector Perturbation}
\end{center}
  \medskip
} \fi

\bigskip
\begin{abstract}
In Statistical Process Control, control charts are often used to detect undesirable behavior of sequentially observed quality characteristics. 
Designing a control chart with desirably low 
False Alarm Rate  (FAR) and detection delay ($ARL_1$) is an important challenge especially when the sampling rate is high and the control chart has an 
In-Control Average Run Length, called $ARL_0$, of 200 or more, as commonly found in practice.
Unfortunately, arbitrary reduction of the FAR typically increases the $ARL_1$. 
Motivated by eigenvector perturbation theory, we propose the Eigenvector Perturbation Control Chart for computationally fast nonparametric profile monitoring. 
Our simulation studies show that it outperforms the competition and achieves both $ARL_1 \approx 1$ and $ARL_0 > 10^6$. 
\end{abstract}

\noindent%
{\it Keywords:}  Statistical Process Control, Nonparametric Profile Monitoring, Change-point Detection, Alarm Fatigue, False Alarm Rate (FAR) 
\vfill

\newpage
\spacingset{1} 
%-----------
%-----------

%-------------------
%-------------------
%-------------------
\section{Overview}
The supplementary materials are organized in the following fashion. 
Section \ref{sec:EigenspaceFixed} derives the eigenspace for the population correlation matrix under the scenario where all profiles are IC and under the scenario where there is a mix of IC and OOC profiles in the window. 
Section \ref{sec:ComputationalConsiderations} discusses computational considerations for the proposed control chart. 
Section \ref{sec:EPCCAlgorithm} details the algorithms involved with the control chart. 
Section \ref{sec:evec_wins_details} covers the details of the simulation study comparing methods with multiple predictors. 
Section \ref{sec:quadquad_details}  discusses details of a  simulation study where quadratic profiles are used for IC and OOC profiles. 
Section \ref{sec:profile_calibration} details the methodology for calibrating profiles to desired effect sizes.  
Section \ref{sec:robotfailures_details} covers a few details regarding the Robot Failures dataset and a note on cleaning the dataset.

%-------------------
%-------------------
%-------------------
\section{Eigenspace of \texorpdfstring{$\bm{\Gamma}$}{$\Gamma$}}\label{sec:EigenspaceFixed}
Assume the are $k_1$ in-control profiles and $k_2$ out-of-control profiles being considered in the sample correlation matrix. 
Let $\gamma_1 $ be the population correlation of two in-control profiles (i.e., $\gamma_1 = (\Cov[f(\bm{x}_i^r),f(\bm{x}_i^s)]+\phi\sigma^2)/(\Var[f(\bm{x}_i^r)]+\sigma^2)$ for $r,s\leq \tau$). 
Similarly, let $\gamma_2$ be the population correlation where two out-of-control profiles are being considered and let $\gamma_{12} = \gamma_{21}$ be the population correlation of an in-control and out-of-control profile.
As the population correlation matrix is going to have a block diagonal structure (aside from the diagonal), let us call $\bm{u}_1 = \sum_{i = 1}^{k_1} \bm{e}_i$ and $\bm{u}_2 = \sum_{i = k_1 + 1}^m \bm{e}_i$ and define $\bm{J}_{ij} = \bm{u}_i \bm{u}_j^\top$ for $i,j\in [2]$. 
Then the population correlation matrix can be written as
\begin{equation}
\bm{\Gamma} = \gamma_{12} \bm{J} + (\gamma_2 - \gamma_{12}) \bm{J}_{22} + (\gamma_1 - \gamma_{12})\bm{J}_{11}  + (1-\gamma_1) \operatorname{diag}(\bm{u}_1) + (1-\gamma_2)\operatorname{diag}(\bm{u}_2)
\end{equation}

As the row sums for the first $k_1$ rows are the same (the same can be said for the last $k_2$ rows), the eigenvector should be of the form $\alpha \bm{u}_1 + \bm{u}_2$. 
By the definition, 
$$\lambda (\alpha \bm{u}_1 + \bm{u}_2)  = \left(\bm{\Gamma} \right) (\alpha \bm{u}_1 + \bm{u}_2) = \left[( 1+ (k_1 - 1)\gamma_1) \alpha + k_2 \gamma_{12}\right] \bm{u}_1 + \left[k_1\gamma_{12} \alpha + (1 + (k_2 -1) \gamma_2)\right] \bm{u}_2.$$
As the coefficients of $\bm{u}_1$ and $\bm{u}_2$ must agree in the left hand and right hand expressions, we can write

\begin{align*}
\begin{bmatrix}
( 1+ (k_1 - 1)\gamma_1) & k_2 \gamma_{12} \\
k_1\gamma_{12} & (1 + (k_2 -1) \gamma_2) 
\end{bmatrix}
\begin{bmatrix}
\alpha \\
1
\end{bmatrix}
= \lambda 
\begin{bmatrix}
\alpha \\
1
\end{bmatrix}
\end{align*}
or equivalently, 
\begin{align*}
\begin{bmatrix}
( 1 - \lambda +(k_1 - 1)\gamma_1) & k_2 \gamma_{12} \\
k_1\gamma_{12} & (1 -\lambda + (k_2 -1) \gamma_2) 
\end{bmatrix}
\begin{bmatrix}
\alpha \\
1
\end{bmatrix}
= 0.
\end{align*}
The second row implies $\lambda = \alpha k_1 \gamma_{12} + 1 + (k_2 - 1) \gamma_2$, and substituting this into the first row gives us
\begin{align*}
k_1 \gamma_{12} \alpha^2 + \left( (k_2-1)\gamma_2 - (k_1 -1) \gamma_1\right) \alpha - k_2 \gamma_{12} = 0.
\end{align*}
The quadratic formula implies $\alpha$ can take on two values 
\begin{align*}
\alpha = \frac{       (k_1 - 1)\gamma_1 - (k_2 - 1)\gamma_2 \pm \sqrt{  \left((k_1 - 1)\gamma_1 - (k_2 - 1)\gamma_2\right)^2 + 4 k_1 k_2 \gamma_{12}^2  }             }{         2 k_1\gamma_{12}            }
\end{align*}
corresponding to the eigenvalues 
\begin{align*}
\lambda = 1+\frac{       (k_1 - 1)\gamma_1 + (k_2 - 1)\gamma_2 \pm \sqrt{  \left((k_1 - 1)\gamma_1 - (k_2 - 1)\gamma_2\right)^2 + 4 k_1 k_2 \gamma_{12}^2  }             }{         2             }.
\end{align*}

%-------------------
%-------------------
%-------------------
\section{Computational Considerations for the eigenvector perturbation control chart}\label{sec:ComputationalConsiderations}

We 
specify the procedure to compute the leading eigenvector of $\bm{R}$.

Power iteration is a simple choice for computing the leading eigenvector of a symmetric matrix. 
A possible concern in using power iteration is the possibility of slow convergence, but we show this concern is negligible in the supplementary materials. 
We claim the dominance of the leading eigenvalue of $\bm{R}$ over the rest of the eigenvalues results in fast convergence.
Power iteration on $\bm{R}$ converges geometrically with ratio $\lambda_2 / \lambda_1$, so slow convergence becomes an issue if $\lambda_2 / \lambda_1\approx 1$ \citep*[sec 5.2]{AcklehNumericalAnalysis}. 
As the eigenvalues of $\bm{R}$ should not deviate greatly from the eigenvalues of $\bm{\Gamma}$, we can use $\tilde{\lambda}_2 / \tilde{\lambda}_1$ to approximate convergence of power iteration on $\bm{R}$. 
It can be shown $\tilde{\lambda}_1 = 1 + \gamma_1 (w - 1)$ and $\tilde{\lambda}_2 = \dots = \tilde{\lambda}_w = 1 - \gamma_1$.
Therefore, 
$ \tilde{\lambda}_2 / \tilde{\lambda}_1 = \left( \frac{ 1 + \gamma_1 ( w - 1)}{1-\gamma_1} \right)^{-1} = \left( \frac{ 1}{1-\gamma_1}+\frac{\gamma_1 }{1-\gamma_1}( w - 1) \right)^{-1}$
which is well separated from 1 if either $w$ is large or $\gamma_1$ is not too small. 
Note $\gamma_1$ is close to zero only if the in-control profile is difficult to distinguish from the noise (e.g., $\Var[f]  \leq \sigma^2$).
We assume the in-control function $f$ is sufficiently distinguishable from the noise, so we conclude that power iteration should converge reasonably quickly.
Leveraging $\bm{v}$ being close to $\frac{1}{\sqrt{w}}\bm{1} $ under in-control conditions, we adopt a modified version of power iteration from \cite*{Iguchi2017} detailed in the Algorithm \ref{alg:PwrIter} which provides an early stopping rule when it becomes evident that the leading eigenvector is not $\frac{1}{\sqrt{w}}\bm{1} $. 
\begin{minipage}{\linewidth}
\begin{lstlisting}[caption={Modified Power Iteration Detector (adapted from \citet{Iguchi2017})},label=alg:PwrIter]
Given:  - Symmetric matrix $\bm{M}\in \mathbb{R}^{w\times w}$
        - Unit eigenvector $\tilde{\bm{v}}\in \mathbb{R}^w$
        - Tolerance: $\zeta>0$
Do: 
Draw $\bm{q}$ uniformly at random from the unit sphere in $\mathbb{R}^w$
while TRUE
	if $|\bm{q}^\top \bm{M} \bm{q}| > |\tilde{\bm{v}}^\top \bm{M} \tilde{\bm{v}}|$ $\,$ ## $\tilde{\bm{v}}$ is not the leading eigenvector of $\bm{M}$
		return($\bm{q}$); break
	else if $(\tilde{\bm{v}}^\top \bm{q})^2 \geq 1 - \zeta$     ## $\tilde{\bm{v}}$ and $\bm{q}$ are close enough
		return($\bm{q}$); break
	end
	$\bm{q} = \bm{Mq} / ||\bm{Mq}||_2$
end
\end{lstlisting}
\end{minipage}

%-------------------
%-------------------
%-------------------
\section{The eigenvector perturbation control chart}\label{sec:EPCCAlgorithm}
Leveraging predictors being fixed for all $t$, we  perform a parametric bootstrap to compute a control limit. 
The procedure for this parametric bootstrap is outlined below for large values of $N$ and $N_0$ with $N_0 >> N$. 
Once a control limit is in hand, the Eigenvector Perturbation Control Chart (Algorithm \ref{alg:eigvCC}) can be executed. 

\begin{enumerate}
\item Compute $\widehat{f}(\bm{X}) = \frac{1}{m} \sum_{t= 1-m}^0 \bm{y}^t$ and
$\hat{\sigma}^2 = \frac{1}{n(m - 1)}\sum_{i=1}^n \sum_{t = 1-m}^0 (\bm{y}^t_i - \widehat{f}(\bm{X})_i)^2$.
\item Compute $\bm{y}^s_\star = \widehat{f}(\bm{X}) + \bm{\epsilon}^s_\star$ for all $s\in [N_0]$ where $\bm{\epsilon}^s_\star \sim N(0, \hat{\sigma}^2 \bm{I})$
\item For every $l\in [N]$, compute a sample correlation matrix $\bm{R}_l$ using $w$ of the $\bm{y}^s_\star$ sampling without replacement. 
\item Compute $S_l = \smash{\displaystyle\max_{k_1\in K}} ||\bm{v}(k_1) - \frac{1}{\sqrt{w}} \bm{1}||_2$ for every bootstrapped correlation matrix $\bm{R}_l$. 
\item Using a small $c>0$, set $U = F(1-c; \hat{\mu}_S, \hat{\sigma}^2_S)$ where $F$ is a normal distribution function for with mean $\hat{\mu}_S$ and variance $\hat{\sigma}^2_S$ and where these estimates are the typical unbiased estimators for the mean and variance of the $\{S_l\}_{l\in [N]}$ in the previous step.   
\end{enumerate}

%-------------------
%-------------------
%-------------------
\section{An illustration of distinct but dependent predictors with the eigenvector perturbation control chart}

\begin{figure}[]
    \centering
    \includegraphics[width = 0.6\linewidth]{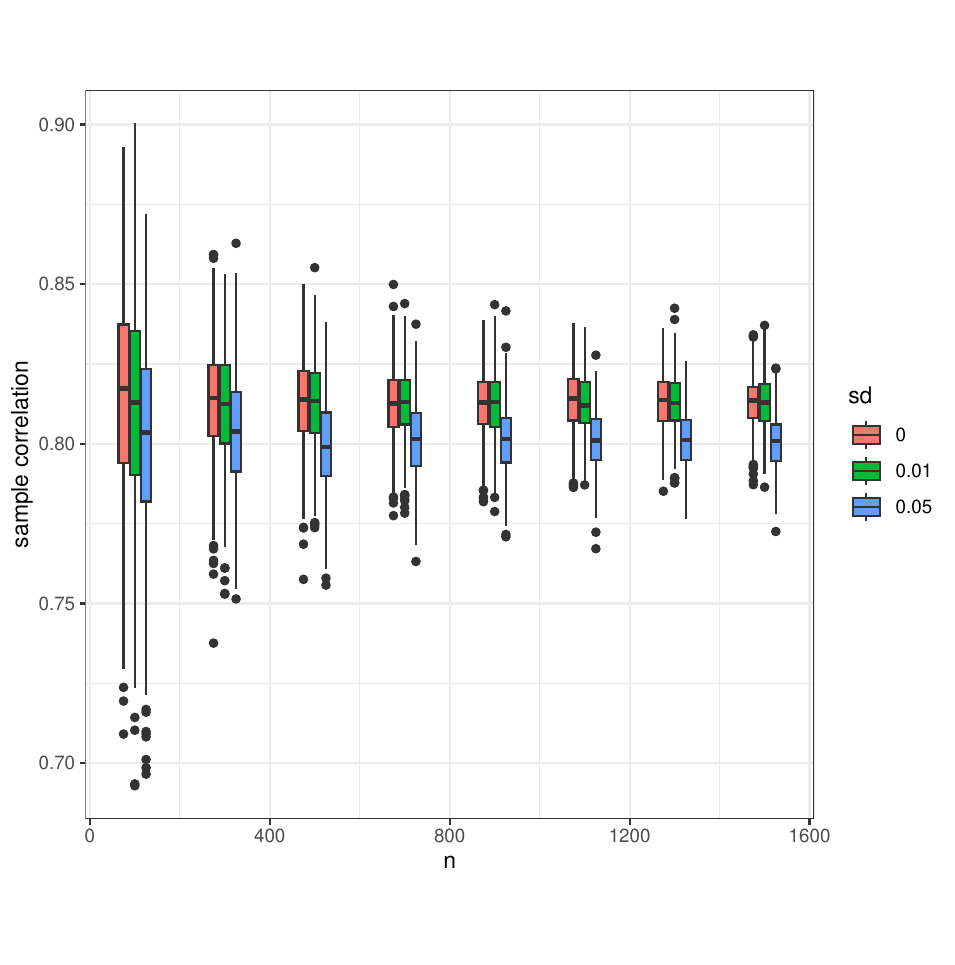}
    \caption{Distribution of sample correlations after perturbation of design points}    \label{fig:cor_peturb_design_points}
\end{figure}

In the main manuscript, we use a common fixed design for the predictors even though we only require the predictors to be dependent between time points (see section 2.1). 
We take a moment to illustrate how this can be relaxed.
To illustrate this, see Figure \ref{fig:cor_peturb_design_points} which is produced from a small simulation study. 
For each value of $n$, 500 trials are conducted wherein each trial a set of $x_1,\dots, x_n$ design points are generated from a uniform(0,1) distribution and then another set of design points $x^*_1, \dots, x^*_n$ are obtained by adding $N(0, \text{sd})$ noise to the design points $x_1, \dots, x_n$.  
The profiles are of the form $y_i = 7\, x_i^2 + \epsilon_i$ for $i \in [n]$ and $\epsilon\sim N(0,1)$. 
We then compute the sample correlation of the responses from the profile using the design points $x_1, \dots, x_n$ with the responses from the profile using the design points $x_1^*, \dots, x_n^*$.
Observe in Figure \ref{fig:cor_peturb_design_points} that although the sample correlations seem to converge to different values, they do converge. 

\begin{figure}[]
\centering
\subfloat[][]{
\includegraphics[width=0.3\linewidth, page = 1]{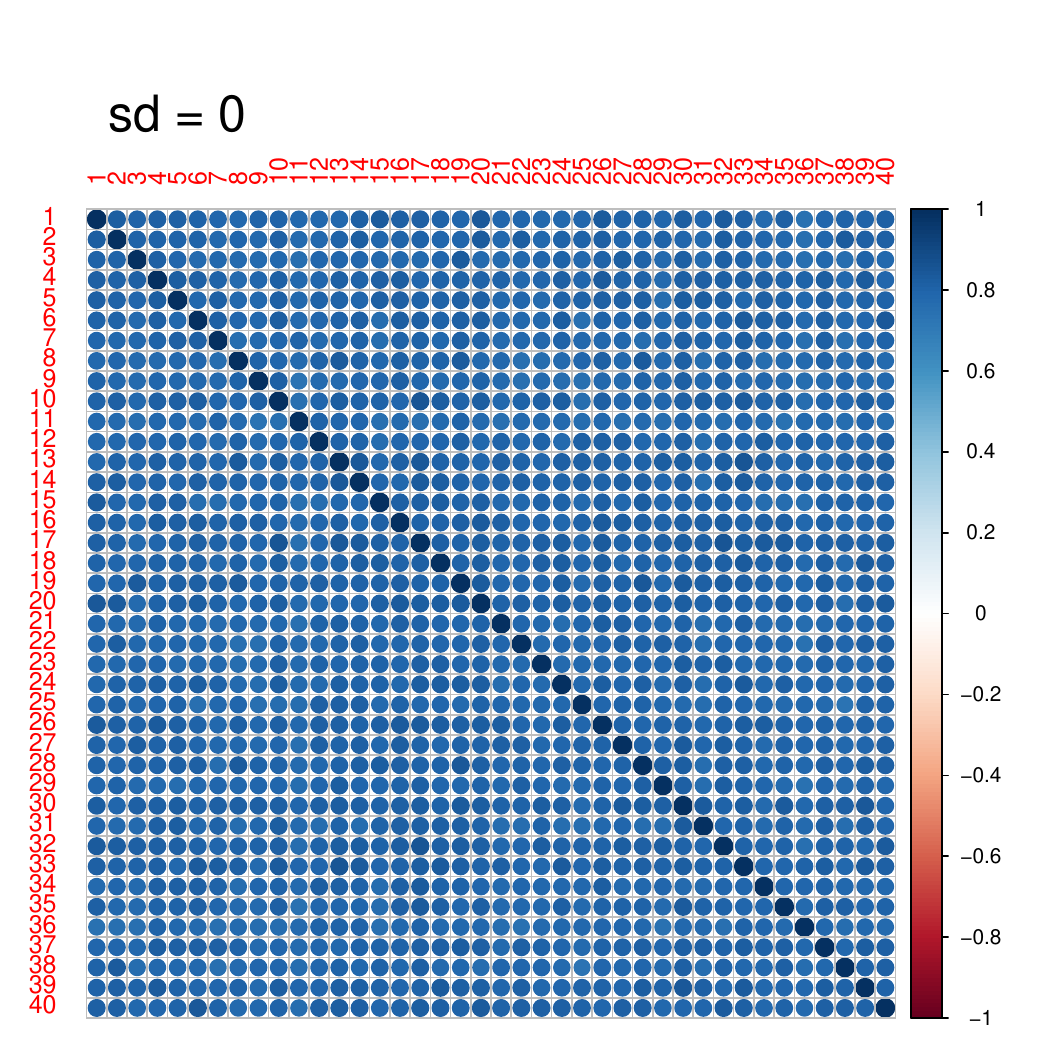}
% \label{fig:visER_allIC}
}
\,\,
\subfloat[][]{
\includegraphics[width=0.3\linewidth, page = 4]{corrplot_petrubed_design_points_varysd.pdf}
% \label{fig:visER_allIC}
}
\,\,
\subfloat[][]{
\includegraphics[width=0.3\linewidth, page = 6]{corrplot_petrubed_design_points_varysd.pdf}
% \label{fig:visER_allIC}
}
\\ % end row 1------
\subfloat[][]{
\includegraphics[width=0.3\linewidth, page = 1]{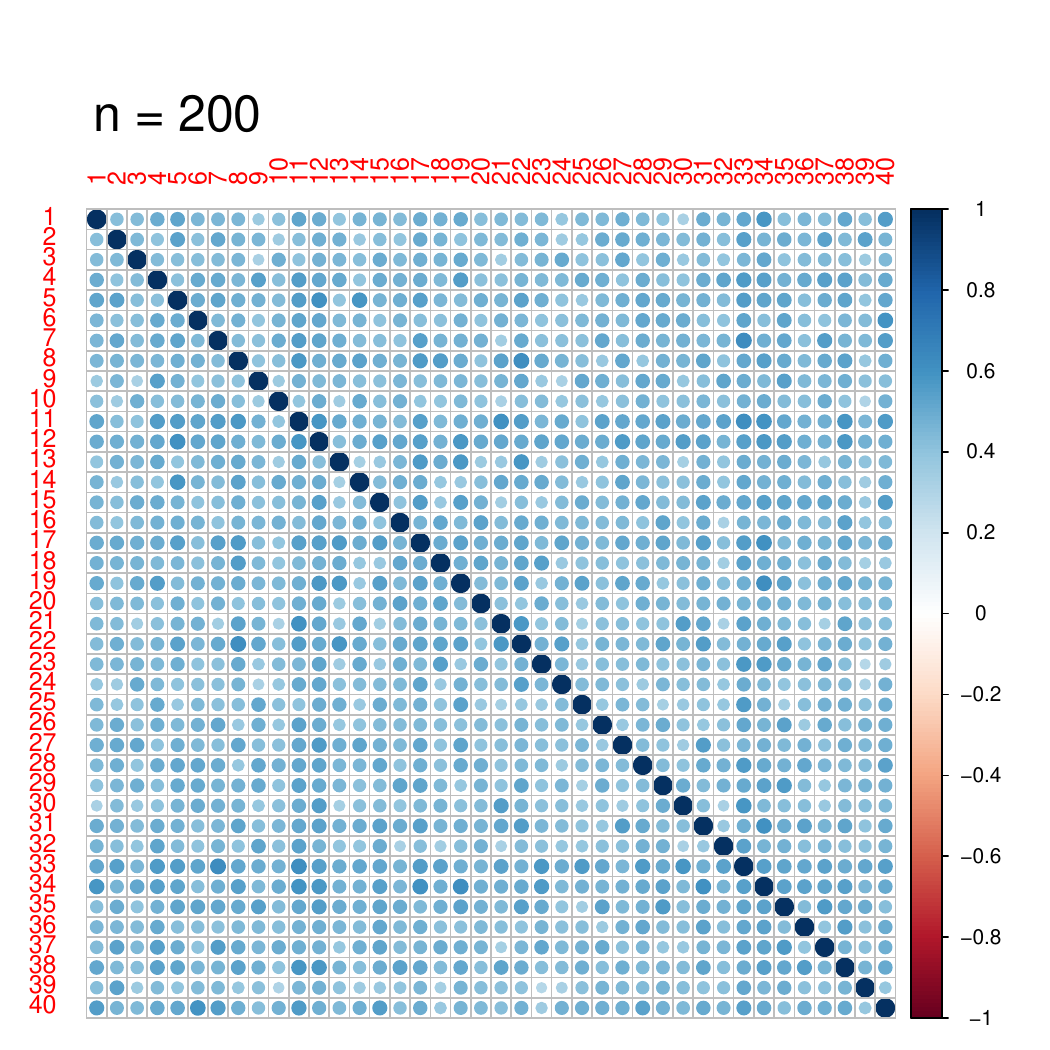}
% \label{fig:visER_allIC}
}
\,\,
\subfloat[][]{
\includegraphics[width=0.3\linewidth, page = 10]{corrplot_petrubed_design_points_varynn.pdf}
% \label{fig:visER_allIC}
}
\,\,
\subfloat[][]{
\includegraphics[width=0.3\linewidth, page = 21]{corrplot_petrubed_design_points_varynn.pdf}
% \label{fig:visER_allIC}
}
\\   % end row 2-------
\subfloat[][]{
\includegraphics[width=0.3\linewidth, page = 1]{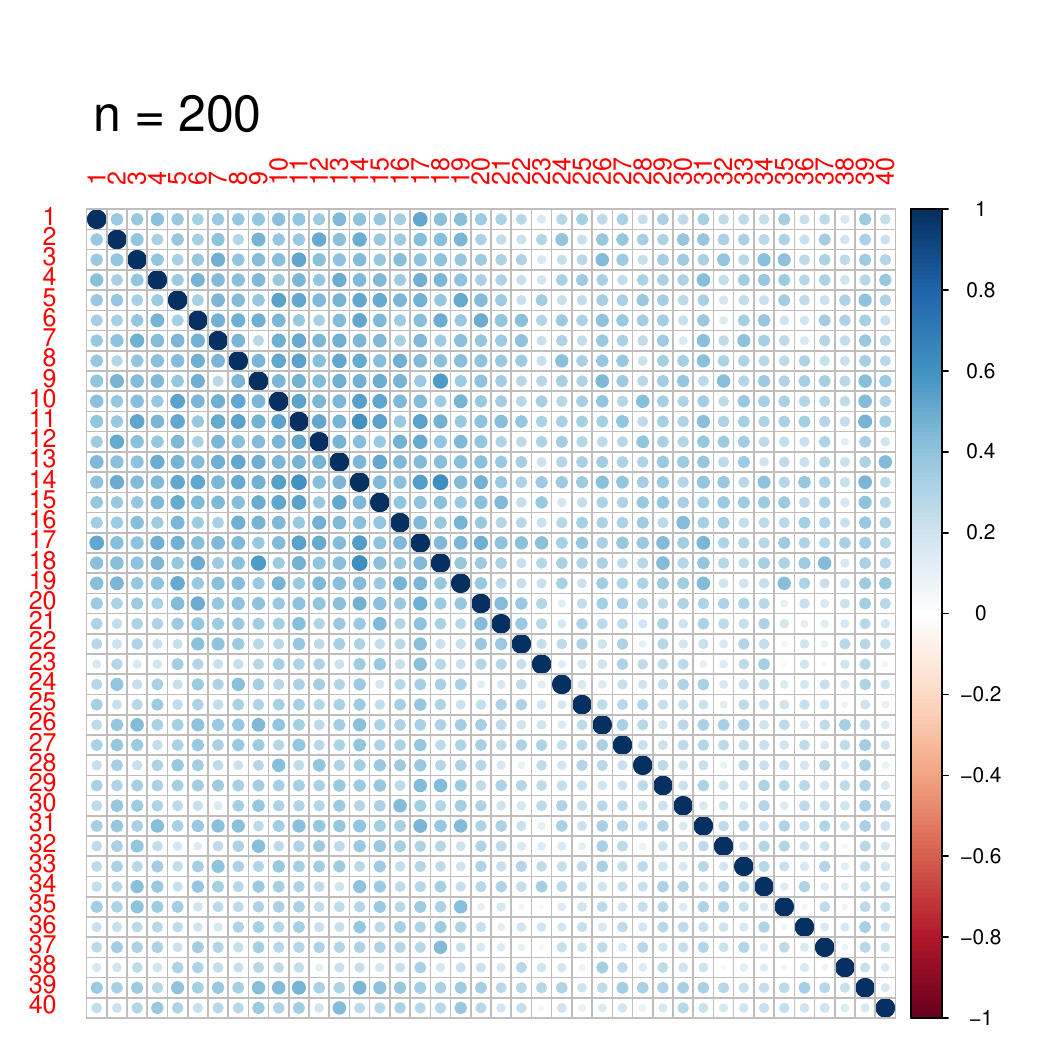}
% \label{fig:visER_allIC}
}
\,\,
\subfloat[][]{
\includegraphics[width=0.3\linewidth, page = 10]{corrplot_petrubed_design_points_varynn_ooc.pdf}
% \label{fig:visER_allIC}
}
\,\,
\subfloat[][]{
\includegraphics[width=0.3\linewidth, page = 21]{corrplot_petrubed_design_points_varynn_ooc.pdf}
% \label{fig:visER_allIC}
}
\caption{
A demonstration of how increasing $n$ can overcome pertubation of design points between profiles. 
(Top row) From left to right the standard deviation of the perturbation increases for IC profiles. 
(Middle row) From left to right the profile sample size $n$ increases and all 40 profiles are IC. 
(Bottom row) From left to right the profile sample size $n$ increases and the first 20 profiles are IC and the last 20 profiles are OOC.  
}
\label{fig:peturbed_design_matrices}
\end{figure}

Ultimately, we care more about the variability of the sample correlation of two profiles. 
The goal of the control chart is to make the appropriate block structure as demonstrated in Figure 1 of the main manuscript as clear as possible. 
If the variability of the sample correlations is large, then it becomes less apparent both to the eye when viewing a heatmap of $\bm{R}$ and to the eigenvector perturbation control chart. 
See the Figure \ref{fig:peturbed_design_matrices} below for a demonstration of this concept. 
In Figure \ref{fig:peturbed_design_matrices}, each sub-figure is obtained by first fixing a set of design points. 
Then 40 observed profiles are generated by first perturbing the design points then inputting the design points into either the in-control function or the out-of-control function and lastly $N(0,1)$ noise is added. 
In the figure above we see as we increase the perturbation of the design points the ``one-block'' structure of the correlation matrix becomes less apparent. 
In the middle row, we fix the perturbation standard deviation to be 0.25 to match Figure 1e. 
Notice the one block structure is regained at the cost of the sample correlations being smaller in magnitude.
The last row uses an out of control profile of $y_i = 3 x_i^2 - x_i + \epsilon_i$. 
Observe using the same values of $n$ as in the middle row allows us to recover a clear two block structure.
Therefore, our control chart can handle design points changing over time so long as they are dependent.

 % \newpage
%-------------------
%-------------------
%-------------------
\section{Details regarding the simulation study comparing the proposed control chart to its competitors with multiple predictors}\label{sec:evec_wins_details}
The factors and levels of the full factorial design of the comparative simulation study are detailed in Table \ref{tbl: factor_level_simest}. 
The values of the convex combination parameter $\nu$ are taken from \citet{Iguchi2021} and are shown in Table \ref{tbl: tgt_w_snr}.
To justify the lower bound of $7 \times 10^6$ for the $ARL_0$ of the proposed control chart for the profile combinations listed in Table \ref{tbl: factor_level_simest}, the proposed control chart was run until either timestep $10^7$ or a false alarm occurred.
The results of carrying out this experiment over 100 trials for each in-control profile yield lower bounds on the $ARL_0$ and are listed in Table \ref{tab: eigv ARL0}. 
Computational speed was also measured across the profile monitoring methods being considered and the results are shown in Table \ref{tab:compare_runtimes}.

The SVR approach from \citet{Li2019} used a smoothing parameter $\lambda$ in their EWMA based scheme and performed a sensitivity analysis for this parameter.
They observed that $ARL_1$ becomes shorter as $\lambda$ becomes smaller.
We used the smallest value ($\lambda=0.05$) from their sensitivity analysis. 
\citet{Li2019} recommended basing the monitoring statistic using the 2nd, 3rd, or 5th \citet{Williams2007} statistics.
As described in \citet{Iguchi2021}, the performance of the SVR approaches on the profile combinations in their simulation study do not differ greatly in $ARL_1$.
As the 5th such statistic resulted in the smallest FAR for the SVR approach in \citet{Iguchi2021}, we also use this statistic for the SVR approach.  

For calibrating control limits using a Monte Carlo approach, control limits were calibrated to a target $ARL_0$ of 200 or 370 using $f$ explicitly to generate profiles. 
Each trial in the Monte Carlo approach consisted of randomly generating a set of historical profiles, fitting a model to the historical profiles, and monitoring a sequence of in-control profiles until a false alarm was raised. 
Control limits were calibrated by finding the smallest observed monitoring statistic that met or exceeded the desired $ARL_0$ over 390 trials.
When the bootstrap approach to setting a control limit is used in the eigenvector perturbation control chart, settings of $\zeta = 10^{-3}$, $c = 10^{-14}$, $N = 1000$, and $N_0 = 5000$ were used.

\begin{table}[h]
\centering
\caption{The factor and levels of the simulation study. The values of $\nu$ were obtained from \cite{Iguchi2021} where they calibrated $\nu \in (0,1)$ to achieve a particular SNR.
}
\scalebox{0.85}{
\begin{tabular}{p{2.9cm}ll}
Factor                      & \multicolumn{2}{c}{Levels} \\ \hline \hline
In-control \&                   & -  $f(\bm{x}) =1+3x_1 + 2x_2 + x_3$ 
                  & $h(\bm{x}) = \nu f(\bm{x}) + (1-\nu) \sin(2\pi x_1 x_2)$\\ 
out-of-control                        & -  $f(\bm{x}) =\frac{4}{9}\left(3x_1 + 2x_2 + x_3\right)^2$ 
                                & $h(\bm{x}) = \nu f(\bm{x}) + (1-\nu) 5 \sin(2\pi x_1 x_2)$ \\ 
functions                            & -  $f(\bm{x}) =1+3x_1 + 2x_2 + x_3$ 
                                & $h(\bm{x}) = \nu f(\bm{x}) + (1-\nu) 25 |x_1-0.5|e^{-x_2}I(x_3 > 0.5)$\\ 
                            & -  $f(\bm{x}) =\frac{4}{9}\left(3x_1 + 2x_2 + x_3\right)^2$ 
                                & $h(\bm{x}) = \nu f(\bm{x}) + (1-\nu) 25 |x_1-0.5|e^{-x_2}I(x_3 > 0.5) $ \\ \hline    
Change point detector       & \multicolumn{2}{l}{
                            \begin{tabular}[l]{@{}l@{}} 
                            - \citet{Li2019} detector with the 5th \citet{Williams2007} statistic\\
                            - \citet{Iguchi2021} \\
                            - Eigenvector Perturbation Control Chart
                            \end{tabular}
                            }\\ \hline
SNR                         & 3, 5    &     \\ \hline
\end{tabular}
}
\label{tbl: factor_level_simest}
\end{table}

\lstset{
    basicstyle=\linespread{0.6}\footnotesize,
    columns=fullflexible,
    breaklines=false,
    mathescape=true,
    %frame=single
    frame=bt
}
\begin{minipage}{0.9\linewidth}
\begin{lstlisting}[caption={Eigenvector Perturbation Control Chart},label={alg:eigvCC}]
Given: $\;\,$- Historical data: $\bm{X}^t,\bm{y}^t$ where $\bm{y}^t\in \mathbb{R}^n,\, \bm{X}^t\in \mathbb{R}^{n \times p} ,\,  t\in \{1-m, \dots, 0\}$
        - Upper control limit: $U>0$
        - A nonempty $K\subset [w-1]$
        - Tolerance for power iteration detector: $\zeta$
        - Window size: $w\leq m$
Do: 
Compute the sample correlation matrix $\bm{R}^\star \in \mathbb{R}^{m \times m}$ of the historical profiles.
Set $\bm{R}$ to be the last $w$ columns and rows of $\bm{R}^\star$
$S = -1$; $T$ = 1
#Conduct process monitoring
$\bf{while}$ $S < U$
    Observe $\bm{y}^T\in \mathbb{R}^{n}$ ; update $\bm{R}$ to reflect correlations of $\{\bm{y}^{T-w+1}, \dots, \bm{y}^{T}\}$
    $\bf{for}$ every $k_1$ in $K$
        $\bf{if}$ $T < w - k_1$, set $\mathcal{I} = [m - w + k_1 + T]$ $\bf{else}$ set $\mathcal{I} = [m]$
        Sample $k_1$ indices $j_1,\dots, j_{k_1}$ from $\mathcal{I}$ without replacement ; set $\bm{R}(k_1)=\bm{R}$
        Replace the first $k_1$ rows/columns of $\bm{R}(k_1)$ with the $j_1,\dots, j_{k_1}$ rows/columns of $\bm{R}^\star$
        $\bm{v}(k_1)= \,\,$modified_power_iteration_detector$\left(\bm{R}(k_1),\frac{1}{\sqrt{w}} \bm{1},  \zeta\right)$
    $S = \max_{k_1\in K} ||\bm{v}(k_1) - \frac{1}{\sqrt{w}} \bm{1}||_2$
    $\bf{if}$ $S > U$, claim change point occurred $\bf{else}$ set $T = T + 1$
\end{lstlisting}
\end{minipage}

\begin{table}[h]
\centering
\caption{Values $v$ for a targeted signal-to-noise ratio}
\begin{tabular}{ccccc}
                                                &                                        & \multicolumn{3}{c}{SNR}                           \\ \cline{3-5} 
$f^0$  & $h$  & 3 & 5 & 7 \\ \hline
\multicolumn{1}{c|}{\multirow{2}{*}{Linear}}    
    & \multicolumn{1}{c|}{Sinusoid}          & \multicolumn{1}{c|}{0.45676033} & \multicolumn{1}{c|}{0.29857462} & 0.16998640  \\ \cline{2-5} 
\multicolumn{1}{c|}{}                           
    & \multicolumn{1}{c|}{Non-differentiable} & \multicolumn{1}{c|}{0.39454966} & \multicolumn{1}{c|}{0.21840187} & 0.07522249\\ \hline
\multicolumn{1}{c|}{\multirow{2}{*}{Nonlinear}} 
    & \multicolumn{1}{c|}{Sinusoid}          & \multicolumn{1}{c|}{0.46154345} & \multicolumn{1}{c|}{0.30484461} & 0.17747209\\ \cline{2-5} 
\multicolumn{1}{c|}{}                           
    & \multicolumn{1}{c|}{Non-differentiable} & \multicolumn{1}{c|}{0.5465315} & \multicolumn{1}{c|}{0.4146351}   & 0.3074320	\\ \hline
\end{tabular}
\label{tbl: tgt_w_snr}
\end{table}
\begin{table}[h]
\centering
\caption{
$ARL_0^*$ of eigenvector perturbation control charts with $10^{-14}$ quantile trick using equally spaced $k_1^\star$ on profiles from Table \ref{tbl: factor_level_simest}.
Due to memory and time constraints, we compute a right censored average run length $ARL_0^*$ taken over runs that raised a false alarm by timestep $10^7$ (and excludes runs having no false alarms by timestep $10^7$).
As the earliest a false alarm could have been raised for the censored trials is $10^7 + 1$, we can place a lower bound on $ARL_0$ by assuming the censored trials all raised an alarm at time step $10^7 + 1$.
}
\begin{tabular}{llrrrrrrrr}
  \hline
In-control profile & $m$ & $ARL_0^*$ & Finished by $T = 10^7$  & Lower bound on $ARL_0 $\\
\hline
quadratic   & 20  & 3947093 & 35 & 7881483\\
linear      & 20  & 2553138 & 37 & 7244662\\
quadratic   & 40  & 4362645 & 29 & 8365168\\
linear      & 40  & 2815431 & 52 & 7068576
\end{tabular}
\label{tab: eigv ARL0}
\end{table}
\begin{table}[h]
    \centering
    \caption{Runtimes in seconds for $100$ iterations of computing the monitoring statistic from an observed profile and deciding if a profile is in-control or out-of-control. 
    The simulations were conducted under $m=20$ historical profiles, a window size of $w=10$ (if applicable), and the profile used for all $100$ iterations was the in-control quadratic profile from Table \ref{tbl: factor_level_simest}, and $m=20$. }
    \begin{tabular}{cccc} 
    \hline
    Method                      & Min       &   Median  & Max       \\ \hline
    Eigenvector Perturbation    & 0.000     &   0.001   & 0.060     \\
    \citet{Li2019}              & 0.021     &   0.025   & 0.041     \\
    \citet{Iguchi2021}          & 15.452    &  28.003   & 41.365    
    \end{tabular}
    \label{tab:compare_runtimes}
\end{table}

%-------------------
%-------------------
%-------------------
\section{Details regarding the simulation study using quadratic in-control and out-of-control profiles}\label{sec:quadquad_details}

\begin{table}[h]
\centering
\caption{The factors and levels for the simulation study on the eigenvector perturbation control chart using quadratic polynomials for the in-control and out-of-control profiles.
All possible treatment combinations were implemented. 
Note certain combinations of $\Var[f]$, SNR, $\rho(f,h)$ and choice of convexity cannot be achieved.}
\begin{tabular}{ll}
% \hline
Factor                             & Level                           \\ \hline \hline
$\tau$                             & 0, 30 , $10^4$                  \\ \hline
$n$                                & 128, 256, 512                   \\ \hline
$m$                                & 20, 40                          \\ \hline
$m/w$                              & 1, 2                            \\ \hline
SNR                                & 3, 5                            \\ \hline
$\Var[f]$                          & 2,4,6                           \\ \hline
$\rho(f,h)$                       & 0.75, 0.9                       \\ \hline
convexity of $h$ with respect to $f$ and $g$ & convex, non-convex combinations \\ \hline
\end{tabular}
\label{tab:quadquadstudy}
\end{table}

If $\nu \in (0,1)$, then $h$ is a convex combination of $f$ and $g$. 
Otherwise, $h$ is a nonconvex combination of $f$ and $g$. 
We additionally constrain $g$ such that $\Cov[f,g] = 0$ to ease profile calibration efforts and improve interpretability of the results. 
Observe if $\nu < 0$, then $\rho(f,h) <0$. 
As we want to consider difficult cases for the eigenvector perturbation control chart we will ignore the scenario where $\nu < 0$.

Recall the eigenvector perturbation of $\bm{\Gamma}$ is a function of  $\gamma_1$, $\gamma_{12}$, and $\gamma_{2}$. 
Although these are biased estimates of their population counterparts, they converge to the population correlation as $n\to \infty$.
As such, we provide an illustration for the population correlations for two in-control profiles $\rho(f+\epsilon_1, f + \epsilon_2)$, for an in-control profiles and an out-of-control profile $\rho(f+\epsilon_1, h + \epsilon_2)$, and for two out-of-control profiles $\rho(h+\epsilon_1, h + \epsilon_2)$ for all possible profile combinations in Table \ref{tab:quadquadstudy}, where $\epsilon_1, \epsilon_2\in \mathbb{R}$ are noise independent of each other and of the predictors $\bm{x}$. 
The quantities are computed as 
\begin{equation*}
    \rho(f+\epsilon_1, f + \epsilon_2) = \frac{\Var[f]}{\Var[f] + \sigma^2}, \quad \rho(h+\epsilon_1, h + \epsilon_2) = \frac{\Var[h]}{\Var[h] + \sigma^2},  
\end{equation*}
and 
\begin{equation*}
\rho(f+\epsilon_1, h + \epsilon_2) = \frac{\Cov[f,h]}{\sqrt{\Var[f] + \sigma^2}\sqrt{\Var[h] + \sigma^2}} = \rho(f,h)\rho(f+\epsilon_1, f + \epsilon_2)^{1/2} \rho(h+\epsilon_1, h + \epsilon_2)^{1/2}
\end{equation*}
where the last equality is obtained through repeated uses of the definition of correlation.

\begin{figure}
    \centering
    \includegraphics[width = 0.8 \textwidth]{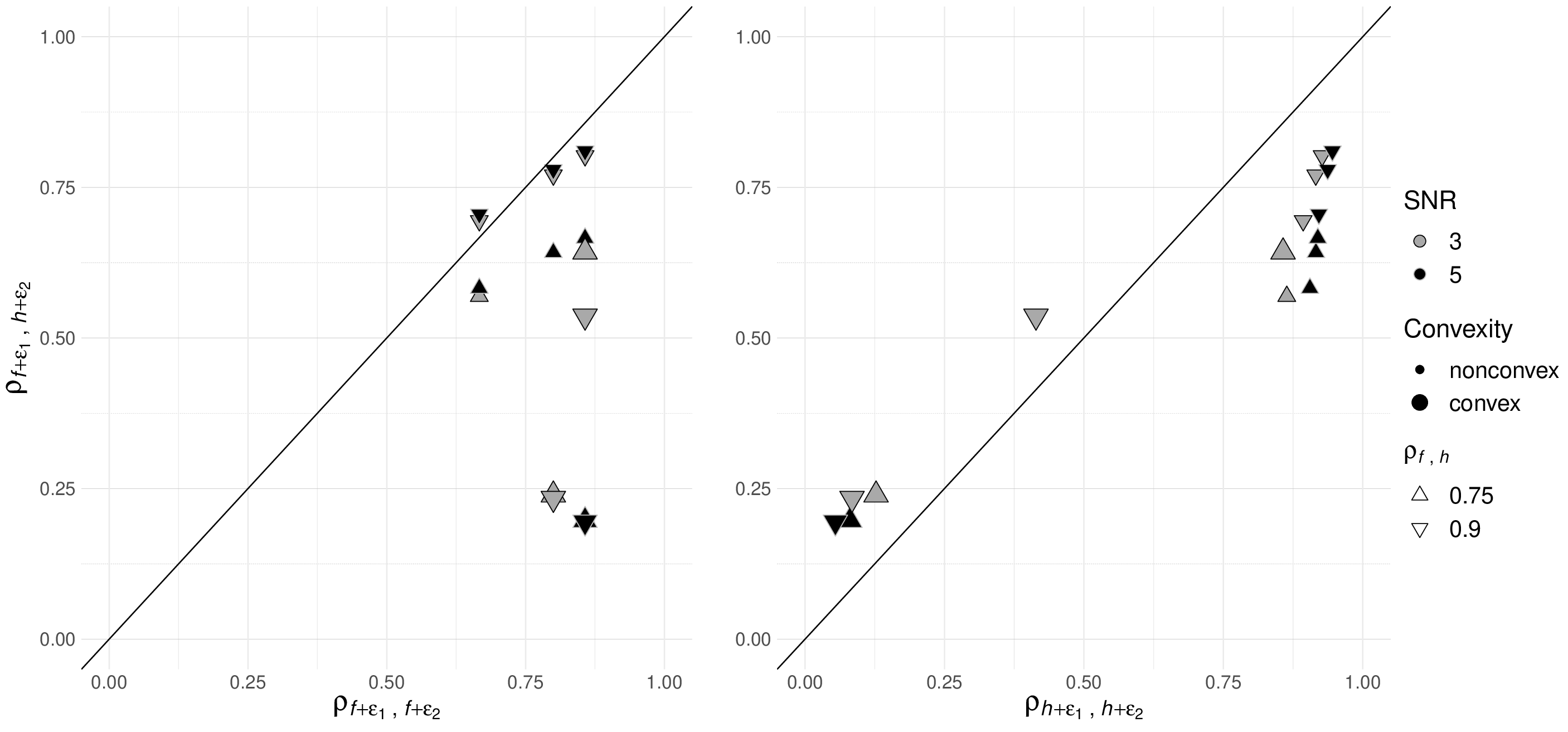}
    \caption{A visualization of the population correlation between profile combinations in Table \ref{tab:quadquadstudy}. 
    Each point corresponds to a particular treatment combination of an in-control profile and an out-of-control profile in Table \ref{tab:quadquadstudy}
    }
    \label{fig:pop_cor_profiles_quadquad}
\end{figure}

There is a counterintuitive result: the $ARL_1$ did not decrease uniformly with increasing SNR and $\Var[f]$ (e.g., $\Var[f] =4)$). 
This curious result is explained by considering the correlation between {\it profiles} as opposed to the correlation between in-control and out-of-control functions $f$ and $h$.
As Figure \ref{fig:pop_cor_profiles_quadquad} shows, the out-of-control profiles obtained through a nonconvex combination of $f$ and $g$ tend to have $\rho(f+\epsilon_1, f+\epsilon_2) \approx \rho(f+\epsilon_1, h+\epsilon_2)$.
In fact, the closer these two quantities are, the larger the $ARL_1$.
It seems the only reason the eigenvector perturbation control chart is able to detect an out-of-control profile is due to the difference between $\rho(h+\epsilon_1, h+\epsilon_2)$ and $\rho(f+\epsilon_1, h+\epsilon_2)$. 
This scenario should be difficult for any control chart that monitors correlations. 

%-------------------
%-------------------
%-------------------
\section{Details regarding competitors omitted from section 4.2.2 }\label{sec:profile_calibration}

We compare two other profile monitoring methods which face computational difficulties in this simulation study. 
\cite{qiu2010NonparProfMonByMixedEffectsModeling}  conjecture their iterative method for estimating a profile does not face convergence issues if both $m$ and $n$ are sufficiently large. 
They recommend values of $n\geq 20$ and $m\geq 500$. 
We attempted to see if this method could still estimate an in-control profile with $m=40$ historical, in-control profiles. 
We attempted estimation 100 times, with each attempt using a different set of $m=40$ profiles, and failed to converge 98 times. 
If their approach fails to estimate the IC function, monitoring cannot occur. 

\begin{figure}[]
    \centering
    \includegraphics[width = 0.5\linewidth]{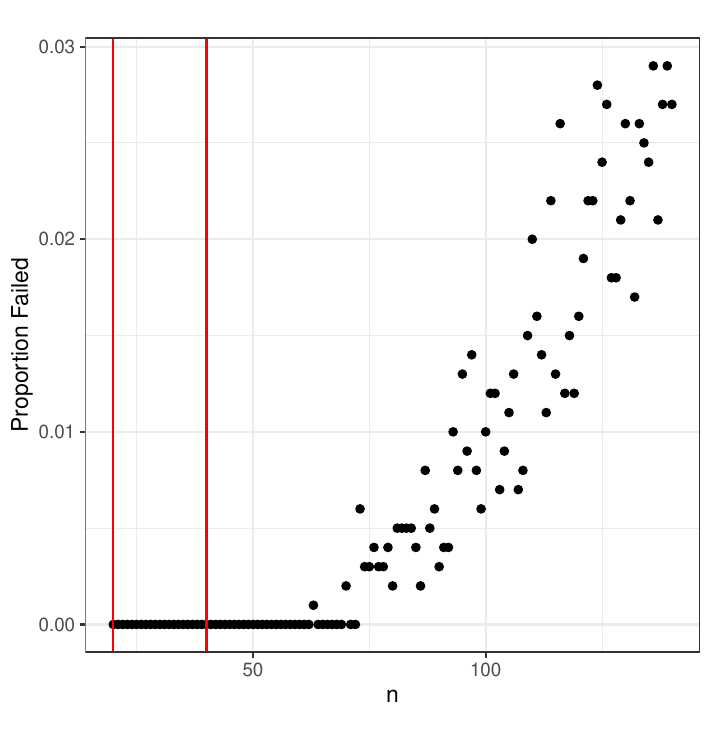}
    \caption{
    Proportion of times the control chart of \cite{Zou2008NonparRegressionProfMon} creates an ill-defined monitoring statistic.
    Red lines correspond to values of $n$ used in the original manuscript.
    }
    \label{fig:Zou2008NonparRegression Fails}
\end{figure}

Similarly, \citet{Zou2008NonparRegressionProfMon} ran into a computational issue for $n > 63$ where we used $n=128$ in section 4.2.2 of the main manuscript.
We were unable to compare the performance of this method in terms of $ARL_1$ and False Alarm Rates with our proposed method, and we provide more detail here than in the main manuscript. 
Their method ran into a computational issue for $n > 63$ where we used $n=128$ in the simulation study in section 4.2.2.
Their approximation of a particular density results in a non-finite estimate of $\sigma^2$ which results in an ill-defined monitoring statistic.  
Specifically, the estimate (using the notation of \cite{Zou2008NonparRegressionProfMon}) $\tilde{\sigma}_j = \Phi^{-1}( \psi(n \hat{\sigma}^2_j))$ is equal to $-\infty$ when $\psi(n \hat{\sigma}^2_j) = 0$. 
The function $\psi$ is the distribution function of the quantity $n \hat{\sigma}^2_j$ which they approximate using the method of Imhoff (1961).
The probability of this approximation incorrectly stating the cumulative probability is zero increases with $n$.  
Figure \ref{fig:Zou2008NonparRegression Fails} demonstrates how often this event occurred as a function of $n$. 
In Figure \ref{fig:Zou2008NonparRegression Fails}, for each value of $n$, $10^3$ trials were performed wherein each trial the predictors were equispaced (see section 3 of Zou et al. (2008)), a profile is randomly generated, and the estimate of $\tilde{\sigma}_j$ was computed. 
This causes an issue as their monitoring statistic requires the multiplication of 0 with $\tilde{\sigma}_j$, which is ill-defined when $\tilde{\sigma}_j = -\infty$. 
We note that this issue was not addressed in Zou et al. (2008) as they used values of $n =20, 40$ in their simulation studies, which are shown in red lines in the above figure.  
So we claim the EP control chart is preferable, provided $n$ is sufficiently large.

%-------------------
%-------------------
%-------------------
\section{A general approach to profile calibration}\label{sec:profile_calibration}
\begin{figure}[ht!]
\centering
\begin{minipage}{.24\textwidth}
  \centering
   \includegraphics[width=\linewidth, page = 1]{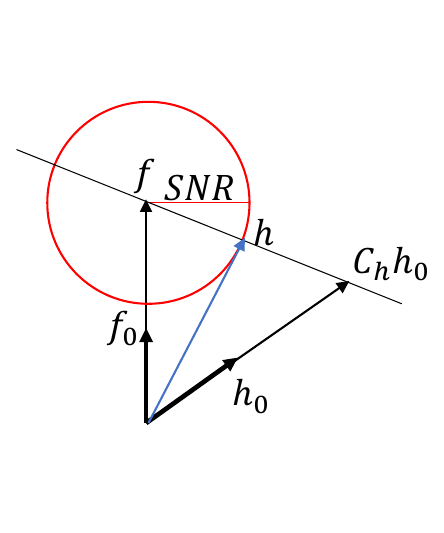}
%  \caption{}
  \label{fig:guess_cal_wo_corr}
\end{minipage}%
\begin{minipage}{.24\textwidth}
  \centering
  \includegraphics[width=\linewidth, page = 2]{visualizing_calibration}
  \label{fig:guess_cal_corr_cone}
\end{minipage}
\begin{minipage}{.24\textwidth}
  \centering
  \includegraphics[width=\linewidth, page = 3]{visualizing_calibration}
  \label{fig:OOCtheta1}
\end{minipage}
\begin{minipage}{.24\textwidth}
  \centering
  \includegraphics[width=\linewidth, page = 4]{visualizing_calibration}
  \label{fig:OOCtheta2}
\end{minipage}
\caption{
Visualizing profile calibration in a functional space. 
(left) Calibrating in-control and out-of-control profiles without the correlation constraint. The $\nu$ parameter is chosen to achieve a desired SNR. 
(middle left) Blue lines correspond to a cone of functions with correlation no less than the desired $\rho(f,h)$.  
Observe adding the correlation constraint requires adjusting $C_h$ in addition to $\nu$.
(middle right)
An out-of-control $h$ denoted by the blue dot is calibrated to a specified the desired $\Var[\delta]$, $\Var[f]$, $\rho(f,h)$.
Also observe in this instance, there are two values of $\nu$ possible. 
The choice of $\nu\in (0,1)$ is shown, where $\nu >1$ would correspond to $h$ lying on the upper left portion of the circle of radius SNR in the figure. 
(right) 
A visualization of the proposed calibration approach for $h$ using orthogonal $f_0$ and $h_0$.  
}
\end{figure}
%-------------------
%-------------------
\subsection{A more general setup}

In profile monitoring, we are provided with an in-control profile $f$ or samples from $f$. 
In a computer experiment, we need to define $f$ first and then define some different out-of-control function $h$. 
Certainly, this could be picked arbitrarily, but we wish to calibrate $h$ to such that certain notions of ``difference''  between $f$ and $h$ are met. 
We retain a signal-to-noise ratio defined as $\sigma^2_{\text{signal}} = \Var[f-h] / \Var[\epsilon]$ as one of such notions. 
There are, however, an infinite number of such $h$ that can meet such a requirement as can be seen by considering a ball of radius $\sigma_{\text{signal}}$ with respect to a norm defined by $||h||_2^2 = \Var[h]$.
We add further notions of difference which are relevant to the eigenvector perturbation control chart. 
If $h\neq f$, then $h$ must be different in some orthogonal direction to $f$. 
Let $h_0$ be a function of unit norm in this orthogonal direction. 
Then we can define $h$ to be an affine combination of $f$ and $C_h h_0$ for some $C_h>0$.
That is, $h = \nu f + (1-\nu) C_h h_0$. 
As the eigenvector perturbation control chart monitors correlations, we will also add the correlation between $f$ and $h$, denoted $\rho(f,h)$ as another constraint. 
For similar reasoning, we add $\Var[f]$ as a constraint on the choice of $f$ as the sample correlation of two observed in-control profiles becomes more stable as $\Var[f] / \Var[\epsilon]$ becomes larger.  
To accommodate this calibration parameter, we pick some function $f_0$ to be a scaled version of $f$ with unit norm. 
In summary, the more general set up for calibrating profiles is below. 
\begin{align} \label{eq:icoocprofdef}
\text{In-control function: } & f = C_f f_0 \nonumber \\
\text{Out-of-control function: } & h = \nu f + (1-\nu)C_h h_0\\
\text{``signal'': } & \delta = f - h = (1-\nu) (C_f f_0 - C_h h_0) \nonumber 
\end{align}
The choices of $f_0, h_0$ can be taken from some random linear combination of basis functions (such that $\Cov[f_0, h_0] = 0$).
Now the task is given $f_0, h_0, \Var[f], \Var[\delta]$, and $\rho(f,h)$, find $\nu, C_f$, and $C_h$. 

Finding $C_f$ follows from the definition of $f_0$: 
$$C_f = \sqrt{\Var[f]}.$$
We introduce the following lemma which  demonstrates ways of finding $\nu$ and $C_h$. 
\begin{lemma}[Identities for computing $\nu$ and $C_h$]
Under the setting in Eq \eqref{eq:icoocprofdef} and assuming $C_h>0$ WLOG, the following identities hold: 
\begin{itemize}
\item ($\nu$ as a function of $C_h$): $\nu = 1 \pm \sqrt{\frac{\Var[\delta]}{\Var[f] + C_h^2 }}\label{eq: ff}$
\item ($C_h$ as a function of $\nu$): $C_h = \sqrt{\frac{\Var[\delta]}{(1-\nu)^2} - \Var[f]}$.
\item ($\rho(f,h)$ and $\nu$ share the same sign): $\rho(f,h) = \frac{\nu}{\sqrt{ 2\nu - 1 + \frac{\Var[\delta]}{\Var[f]} }}$
\item ($\nu$ as a function of $\rho(f,h)$): $\nu = \rho(f,h)^2 \pm \sqrt{\rho(f,h)^4 + \rho(f,h)^2\left(\frac{\Var[\delta]}{\Var[f]} - 1\right)}$.
\end{itemize}
\end{lemma}
\begin{proof}
As $\Var[\delta] = (1-\nu)^2 \Var[f - C_h h_0]$, and $\Cov[f, h_0] = 0$ by design, 
$$\Var[\delta] = (1-\nu)^2 (\Var[f] + C_h^2 \Var[h_0]), $$
and 
\begin{equation}\label{eq: 1_nu2}
(1-\nu)^2 = \frac{\Var[\delta]}{\Var[f] + C_h^2 \Var[h_0]}.
\end{equation}
Therefore, 
\begin{equation}\label{eq: nuC_h}
\nu = 1 \pm \sqrt{\frac{\Var[\delta]}{\Var[f] + C_h^2 \Var[h_0]}}
\end{equation}
which is solely a function of $C_h$. 
By the definition of correlation and the construction of orthogonal $f_0$ and $h_0$, 
\begin{align*}
\rho(f,h)
& = \frac{\Cov[f,h]}{\sqrt{\Var[f] \Var[h]}}\\
& = \frac{\Cov[f,\nu f + (1-\nu) C_h h_0]}{\sqrt{\Var[f] \Var[h]}}\\
& = \frac{\nu \Var[f] + (1-\nu)C_h \Cov[f,h_0]}{\sqrt{\Var[f] \Var[h]}}\\
& = \frac{\nu \Var[f]}{\sqrt{\Var[f] \Var[h]}}\\
& = \nu \sqrt{\frac{\Var[f]}{ \Var[h]}}\\
& = \nu \sqrt{\frac{\Var[f]}{ \nu^2 \Var[f] + (1-\nu)^2 C_h^2 \Var[h_0]}}\\
& =  \frac{\nu}{ \sqrt{\nu^2 + (1-\nu)^2 C_h^2 \frac{\Var[h_0]}{\Var[f]}}}.
\end{align*}
Rewriting Eq. \eqref{eq: 1_nu2} as 
\begin{equation*}
C_h^2 = \frac{1}{\Var[h_0]} \left( \frac{\Var[\delta]}{(1-\nu)^2} - \Var[f]\right). 
\end{equation*}
Substituting this expression into the last formulation of $\rho(f,h)$ gives
\begin{align}\label{eq:rho_nu_same_sign}
\rho(f,h) 
 = \frac{\nu}{\sqrt{ \nu^2 + \frac{\Var[\delta]}{\Var[f]} - (1-\nu^2)}}
 = \frac{\nu}{\sqrt{ 2\nu - 1 + \frac{\Var[\delta]}{\Var[f]} }}. 
\end{align}
We can rewrite Eq. \eqref{eq:rho_nu_same_sign} as a quadratic in $\nu$ to obtain an expression of $\nu$ in terms of $\rho(f,h)$: 
$$\nu^2 - 2 \rho(f,h)^2 + \rho(f,h)^2 \left(1 - \frac{\Var[\delta]}{\Var[f]}\right).$$
The roots of the quadratic after some simplification are 
\begin{align}\label{eq:nu_rho}
\nu = \rho(f,h)^2 \pm \sqrt{\rho(f,h)^4 + \rho(f,h)^2\left(\frac{\Var[\delta]}{\Var[f]} - 1\right)}.
\end{align}
\end{proof}

\begin{figure}[ht!]
\centering
\begin{minipage}{.33\textwidth}
  \centering
   \includegraphics[width=\linewidth, page = 1]{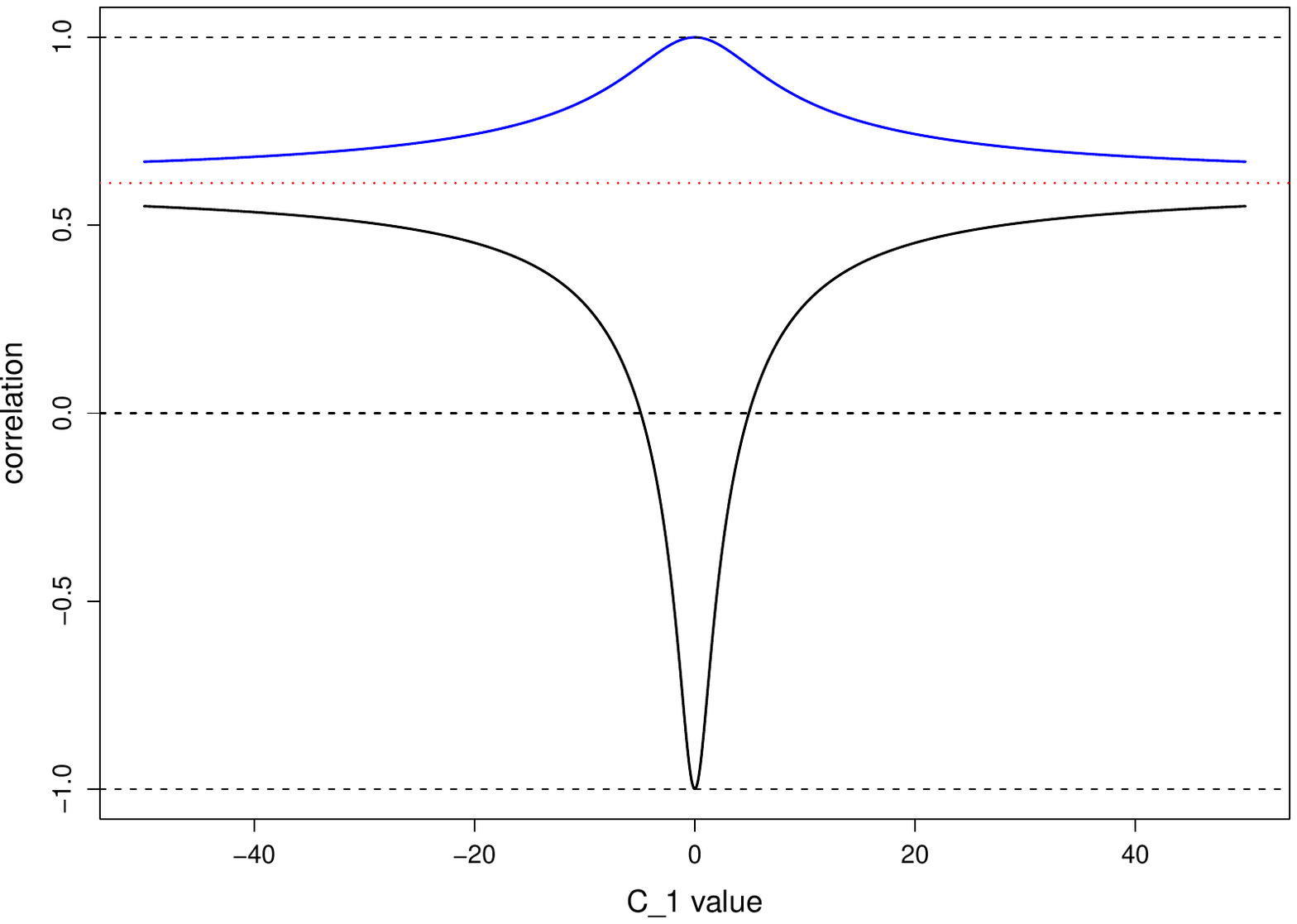}
\end{minipage}%
\begin{minipage}{.33\textwidth}
  \centering
  \includegraphics[width=\linewidth, page = 1]{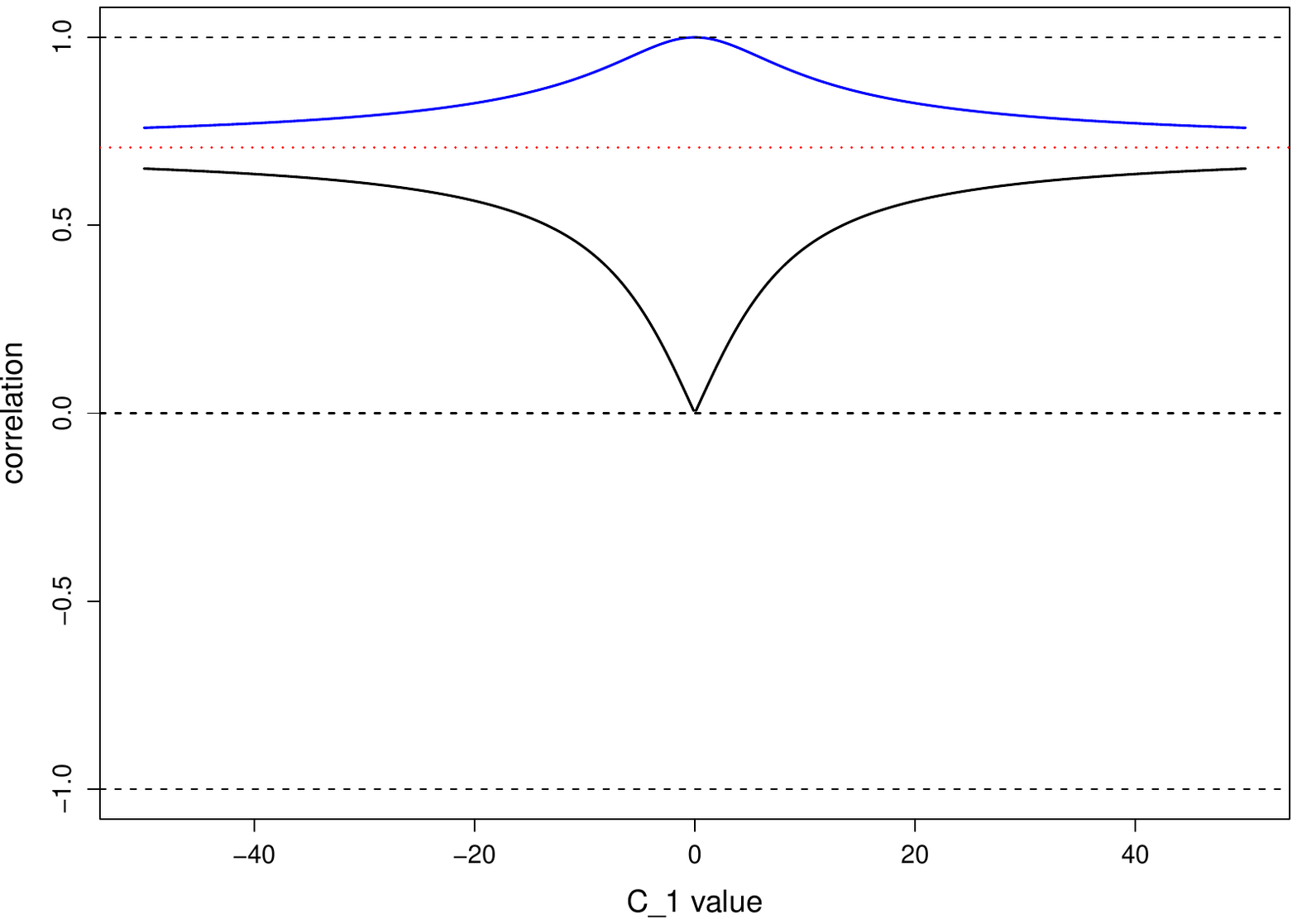}
\end{minipage}
\begin{minipage}{.33\textwidth}
  \centering
  \includegraphics[width=\linewidth, page = 1]{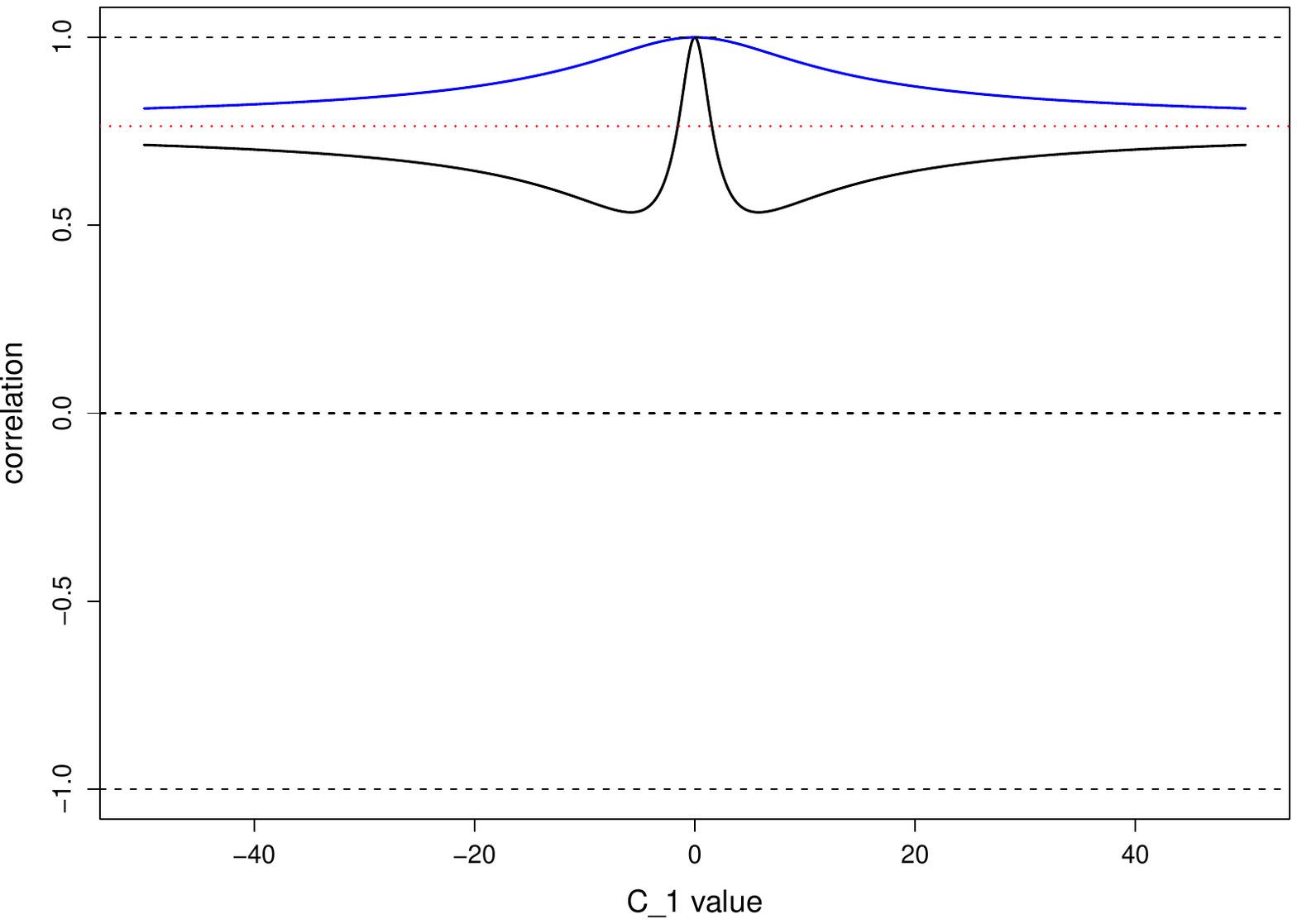}
\end{minipage}
\caption{
A plot of $\rho(f,h)$ as a function of $C_h$. 
The top blue line corresponds to the positive root in the expression for $\nu$ in Eq. \eqref{eq: nuC_h}. 
The bottom black line corresponds to the negative root. 
The horizontal red dotted line corresponds to the horizontal asymptote of $\rho(f,h)(C_h)$. 
In each plot, $\Var[\delta] = 5$. 
Moving from left to right, the values of $\Var[f]$ are 3,5, and 7. 
}
\label{fig:cor_c1_cases}
\end{figure}

Note that there may be two choices of $\nu, C_h$, which can achieve the same calibration. 
It turns out the behavior of $\rho(f,h)$ as a function of $C_h$ can be classified into three cases illustrated in Figure \ref{fig:cor_c1_cases}. 
In the cases where $\Var[f] \leq \Var[\delta]$, the horizontal asymptote (which happens to be $1/\sqrt{1+\Var[\delta]/\Var[f]}$) dictates which root to use to achieve a desired correlation. 
In the case where $\Var[f] > \Var[\delta]$, correlations above the asymptote can be achieved using either root in Eq. \eqref{eq: nuC_h}. 
Additionally, observe when $\Var[f] \geq \Var[\delta]$ there and there exists a minimum achievable correlation. 
As a result of the discriminant in \eqref{eq:nu_rho} being positive, the minimum achievable correlation when $\Var[f] > \Var[\delta]$ is $\sqrt{1 - \frac{\Var[\delta]}{\Var[f]}}$.

%-------------------
%-------------------
\subsection{Calibrating for linear and quadratic polynomial profiles}
Here we assume $f_0$ and $h_0$ are of the following form: 
\begin{align*}
f_0({\bf x}) & = {\bf x}^\top \bm{A} {\bf x} + \bm{a}^\top {\bf x} \\
h_0({\bf x}) & = {\bf x}^\top \bm{B} {\bf x} + \bm{b}^\top {\bf x} 
\end{align*}
We can randomly pick $f_0$ but the choice of $h_0$ will depend on the choice of $f_0$ (in that $\Cov[f_0, h_0] = 0$). 
Additionally, normalizing a random choice of quadratic polynomial to obtain $f_0$ will require us to compute $\Var[f_0]$. 
In order to quickly calibrate profiles, we will need a fast computation of the covariance. 
The covariance between $f_0$ and $h_0$ can be written as
\begin{align}
\Cov[f_0, h_0]
& = \Cov[{\bf x}^\top \bm{A} {\bf x} + \bm{a}^\top {\bf x}, {\bf x}^\top \bm{B} {\bf x} + \bm{b}^\top {\bf x}  ] \nonumber \\
& = 	\Cov[{\bf x}^\top \bm{A} {\bf x}, {\bf x}^\top \bm{B} {\bf x}] +
		\Cov[{\bf x}^\top \bm{A} {\bf x}, \bm{b}^\top {\bf x}] +
		\Cov[a^\top {\bf x}, {\bf x}^\top \bm{B} {\bf x}] +	
		\Cov[\bm{a}^\top  {\bf x}, \bm{b}^\top {\bf x}]. \label{eq:snrcal_covdecomp}
\end{align} 
We handle each term in turn, in reverse order. 
The last term can be written as 
$$\Cov[\bm{a}^\top {\bf x}, \bm{b}^\top {\bf x}] = \bm{a}^\top \bm{b} / 12$$
and the middle terms can be written 
\begin{align*}
\Cov[{\bf x}^\top \bm{A} {\bf x}, \bm{b}^\top {\bf x}]
& = \frac{1}{24} 1^\top (\bm{A} + \bm{A}^\top) \bm{b} \quad \text{and}\\
\Cov[\bm{a}^\top {\bf x}, {\bf x}^\top \bm{B} {\bf x}] 
& = \frac{1}{24} 1^\top (\bm{B} + \bm{B}^\top) \bm{a}.
\end{align*}
WLOG we show the first identity with the following calculation:
\begin{align*}
\Cov[{\bf x}^\top \bm{A} {\bf x}, \bm{a}^\top {\bf x}]
& = \Cov\left(\sum_{i,j} A_{ij} X_i X_j, \sum_k a_k X_k\right)\\
& = \sum_{i,j,k} A_{ij} a_k \Cov[X_i X_j, X_k]\\
& = \sum_{i = k \neq j} A_{ij} a_k \E[ X_j ]\Var[X_i] + \sum_{j = k \neq i} A_{ij} a_k \E [X_j] \Var[X_i] + \sum_i A_{ii} a_i \Cov[X_i^2, X_i]\\
& = \frac{1}{24} \sum_{i = k \neq j} A_{ij} a_k + \frac{1}{24} \sum_{j = k \neq i} A_{ij} a_k + \frac{1}{12} \sum_i A_{ii} a_i \\
& = \frac{1}{24} \left( \sum_{ij} A_{ij} a_i + \sum_{ij} A_{ij} a_j\right)\\
& = \frac{1}{24} \bm{1}^\top (\bm{A} + \bm{A}^\top) \bm{a}.
\end{align*}
The first term in \eqref{eq:snrcal_covdecomp} is rather complicated and is the reason why calculating $\Var (f)$ is slow for a quadratic profile. 
The lemma below gives the expression for this term. 
\begin{lemma}
Let ${\bf x} = (X_1, \dots, X_d)^\top \sim \operatorname{Unif}\left([0,1]^d\right)$ with independent components. Then for $\bm{A},\bm{B}$ defined above, 
\begin{align}\label{eq: covtwoquadraticforms}
\Cov[{\bf x}^\top \bm{A} {\bf x}, {\bf x}^\top \bm{B} {\bf x}] \nonumber
& = \E [{\bf x}^\top \bm{A} {\bf x} {\bf x}^\top \bm{B} {\bf x}] - \E [{\bf x}^\top \bm{A} {\bf x}] \E[ {\bf x}^\top \bm{B} {\bf x}] \\ \nonumber
& = \frac{1}{2^4} \sum_{i,j,k,l\text{ distinct}}  A_{ij}B_{kl} 
+ \frac{1}{6} \sum_{\text{3 distinct }i,j,k,l}  A_{ij}B_{kl} 
 + \frac{1}{3^2} \sum_{\substack{\text{2 distinct }i,j,k,l \\ \text{ each of size 2}}}  A_{ii}B_{kl}  \\ \nonumber
& + \frac{1}{2 \cdot 4} \sum_{\substack{\text{2 distinct }i,j,k,l \\ \text{ one of size 1 the other of size 3}}}  A_{ii}B_{kl}  
+ \frac{1}{5} \sum_{i}  A_{ii}B_{ii} \\ 
& - \left(
\frac{1}{12}\operatorname{tr}(\bm{A}) + \frac{1}{4} \bm{1}^\top \bm{A} \bm{1}
\right) 
\left(
\frac{1}{12}\operatorname{tr}(\bm{B}) + \frac{1}{4} \bm{1}^\top \bm{B} \bm{1}
\right). 
\end{align}
\end{lemma}
\begin{proof}
By a well known identity, 
\begin{align*}
\Cov[{\bf x}^\top \bm{A}{\bf x}, {\bf x}^\top \bm{B}{\bf x}]
& = \E[{\bf x}^\top \bm{A} {\bf x} {\bf x}^\top \bm{B}{\bf x}] - \E[{\bf x}^\top \bm{A}{\bf x}] \E [{\bf x}^\top \bm{B}{\bf x}].
\end{align*}
By taking the bilinear form of ${\bf x}^\top \bm{A}{\bf x}$ we can compute its expectation. 
\begin{align*}
\E [{\bf x}^\top \bm{A}{\bf x}] 
& = \E \left[\sum_{i,j} A_{ij} X_i X_j\right]\\
& = \E \left[\sum_{i,j} A_{ij} X_i X_j\right]\\
& = \sum_i A_{ii} \E \left[X_i^2 + \sum_{i\neq j} A_{ij}\right] \E [X_i] \E [X_j]\\
& = \sum_i A_{ii} \left(\frac{1}{12} + \frac{1}{4}\right) + \sum_{i\neq j} A_{ij}\frac{1}{4} \\
& = \frac{1}{12} \operatorname{tr}(\bm{A}) + \frac{1}{4} \bm{1}^\top \bm{A} \bm{1}
\end{align*} 
A similar result holds for $\bm{B}$. 
Now we need an expression for $ \E [{\bf x}^\top \bm{A} {\bf x} {\bf x}^\top \bm{B}{\bf x}]$. 
Using the fact that ${\bf x}^n \sim \operatorname{Beta}(1/n, 1)$, we can write
\begin{align*}
\E [{\bf x}^\top \bm{A} {\bf x} {\bf x}^\top \bm{B}{\bf x}] 
& = \E \left[\left(\sum_{i,j} X_i A_{ij} X_j\right)\left(\sum_{k,l} X_k B_{kl} X_l\right) \right]\\
& = \sum_{i,j,k,l}  A_{ij}B_{kl} \E [X_i X_j X_k X_l] \\
& = \frac{1}{2^4} \sum_{i,j,k,l\text{ distinct}}  A_{ij}B_{kl} 
+ \frac{1}{6} \sum_{\text{3 distinct }i,j,k,l}  A_{ij}B_{kl} 
 + \frac{1}{3^2} \sum_{\substack{\text{2 distinct }i,j,k,l \\ \text{ each of size 2}}}  A_{ii}B_{kl}  \\
& + \frac{1}{2 \cdot 4} \sum_{\substack{\text{2 distinct }i,j,k,l \\ \text{ one of size 1 the other of size 3}}}  A_{ii}B_{kl}  
+ \frac{1}{5} \sum_{i}  A_{ii}B_{ii}. 
\end{align*}
\end{proof}
A na\"ive way to do calculate $\E [{\bf x}^\top A {\bf x} {\bf x}^\top \bm{B} {\bf x}]$ is to iterate through all possible combinations of $i,j,k,l$ and compute the number of unique indices.
As calling the \texttt{unique} command on the indices is slow, we can reformulate Eq \eqref{eq: covtwoquadraticforms} with the use of entrywise multiplication. 
Define a matrix $\bm{C}(i,j)$ as a function of $i,j$ whose entries correspond to the coefficients of the sums in \eqref{eq: covtwoquadraticforms} where the indices of $\bm{C}(i,j)$ correspond to the indices of $\bm{B}$. 
An example of the $5 \times 5$ case where $i = 2$ and $j=4$ is given by
\begin{equation*}
\bm{C}(i, j) = 
  \left(\begin{array}{c>{\columncolor{black!10}}cc>{\columncolor{black!10}}ccc}
    \mycellcolorlight \nicefrac{1}{12}  & \nicefrac{1}{12}  & \nicefrac{1}{16} & \nicefrac{1}{12} & \nicefrac{1}{16}\\
    \rowcolor{black!10}
    \nicefrac{1}{12}   & \mycellcolordarker \nicefrac{1}{8}  & \nicefrac{1}{12} & \mycellcolordark \nicefrac{1}{9} & \nicefrac{1}{12} \\
    \nicefrac{1}{16}  & \nicefrac{1}{12}  & \mycellcolorlight \nicefrac{1}{12} & \nicefrac{1}{12} &\nicefrac{1}{16}\\
   \rowcolor{black!10}
    \nicefrac{1}{12}   & \mycellcolordark \nicefrac{1}{9}   & \nicefrac{1}{12} & \mycellcolordarker \nicefrac{1}{8} & \nicefrac{1}{12} \\
    \nicefrac{1}{16}   & \nicefrac{1}{12}   & \nicefrac{1}{16} & \nicefrac{1}{12} &  \mycellcolorlight  \nicefrac{1}{12} \\
  \end{array}\right),
\end{equation*}
and the case where $i = j = 2$ is given by
\begin{equation*}
\bm{C}(i, j) = 
  \left(\begin{array}{>{\columncolor{black!10}}c>{\columncolor{black!50}}c>{\columncolor{black!10}}c>{\columncolor{black!10}}c>{\columncolor{black!10}}c}
    \mycellcolordark \nicefrac{1}{9}  & \nicefrac{1}{8}  & \nicefrac{1}{12} & \nicefrac{1}{12} & \nicefrac{1}{12} \\
     \rowcolor{black!50}
    \nicefrac{1}{8}   & \mycellcolordarkest {\color{white}\nicefrac{1}{5}}  & \nicefrac{1}{8} & \nicefrac{1}{8} & \nicefrac{1}{8}\\
    \nicefrac{1}{12}   & \nicefrac{1}{8}   & \mycellcolordark \nicefrac{1}{9} & \nicefrac{1}{12} & \nicefrac{1}{12}\\
    \nicefrac{1}{12}   & \nicefrac{1}{8}   & \nicefrac{1}{12}  & \mycellcolordark \nicefrac{1}{9} & \nicefrac{1}{12}\\
   \nicefrac{1}{12}  & \nicefrac{1}{8}  & \nicefrac{1}{12} & \nicefrac{1}{12} & \mycellcolordark \nicefrac{1}{9}\\
  \end{array}\right).
\end{equation*}
If $\odot$ denotes the entrywise product of two matrices, we can express \eqref{eq: covtwoquadraticforms} as
\begin{align*}
\Cov[{\bf x}^\top A {\bf x} , {\bf x}^\top \bm{B} {\bf x}] 
& = \bm{1}^\top \left( \sum_{i,j} A_{ij} \left(\bm{B} \odot \bm{C}(i,j)\right)\right) \bm{1}\\
& \qquad - \left(
\frac{1}{12}\operatorname{tr}(\bm{A}) + \frac{1}{4} \bm{1}^\top \bm{A} \bm{1}
\right) 
\left(
\frac{1}{12}\operatorname{tr}(\bm{B}) + \frac{1}{4} \bm{1}^\top \bm{B} \bm{1}
\right).
\end{align*}

Now that the computation of the variance and covariance of quadratic polynomials has been covered, we can normalize randomly selected polynomials to obtain a choice of $f_0$. 
Now, we need a way to find an orthogonal $h_0$ to a randomly selected $f_0$.

For the space of polynomials with zero intercept, $\sqrt{\Var[f]}$ is a norm. 
As such, we can use one step of a modified Gram-Schmidt procedure to obtain an orthogonal polynomial $h_0$ to a randomly selected polynomial $f_0$ for purposes of SNR calibration. 
% For purposes of completeness, below is a proof for $\sqrt{\Var(f)}$ being a norm. 
% \begin{proof}
% \begin{itemize}
% \item[(Triangle Inequality)] By H\"{o}lder's inequality, $\Cov (f,g)^2 \leq \Var(f) \Var(g)$. 
% By taking the square root and adding some terms, $\Var(f) + \Var(g) + 2 \Cov(f,g) \leq  \Var(f) + 2\sqrt{\Var(f)\Var(g)} + \Var(g) = (\sqrt{\Var(f)} + \sqrt{\Var(g)})^2$. 
% Taking the square root again yields $$\sqrt{\Var(f)} + \sqrt{\Var(g)} \geq \sqrt{\Var(f) + \Var(g) + 2\Cov(f,g)}  = \sqrt{\Var(f+g)},$$
% which gives the desired result.
% \item[(Absolute homogeneity)] Let $s\in \mathbb{R}$. Then $\sqrt{\Var(sf)} = |s| \sqrt{\Var(f)}$. 
% \item[(Positive definiteness)] Let $f$ be a polynomial with zero intercept such that $\sqrt{\Var(f)}=0$. Then $\Var(f) = 0$ and $f$ must be a constant function. As $f(0) = 0$, $f$ is the zero function.
% \end{itemize}
% \end{proof}
Now assume $f_0$ is randomly generated to have unit norm. 
Let $h_0^*$ also be a randomly generated polynomial with unit norm. 
Using one step of a modified Gram-Schmidt procedure we set $h_0$ to be the following  polynomial:
$$h_0 := h_0^* - \frac{\langle f_0, h_0^*\rangle}{||f_0||}\frac{f_0}{||f_0||}.$$
With this construction, $h_0 \perp f_0$ as 
\begin{align*}
\langle f_0 , h_0 \rangle 
&= \left\langle f_0,  h_0^* - \frac{\langle f_0, h_0^*\rangle}{||f_0||}\frac{f_0}{||f_0||} \right\rangle \\
&= \langle f_0, h_0^*\rangle - \left\langle f_0, \frac{\langle f_0, h_0^*\rangle}{||f_0||}\frac{f_0}{||f_0||} \right\rangle \\
&= \langle f_0, h_0^* \rangle - \langle f_0, h_0^*\rangle \frac{\langle f_0, f_0 \rangle}{||f_0||^2}\\
&=0.
\end{align*}
%-------------------
%-------------------
\subsection{ Example: Univariate predictor profiles in section 4.2.2 of the main manuscript }
The functions used in section 4.2.2 are below:
\begin{align*}
    f_0(x) 
    & = 0.14953379 \, x^2 - 0.09938939 \, x - 0.32320738\\
    h_0^*(x)  
    & = 0.42598510 \,  x^2 - 0.17591160 \, x + 0.47024360.
\end{align*}
The calibrated profiles are below for when $\Var[f] = 6, \Var[\delta], \rho(f,h) = 0.9$ and a convex combination is used:
\begin{align*}
    f(x) 
    & = 134.0778 \, f_0(x)\\
    h(x)  
    & = 35.02044 \, f_0(x)  -5.536481\, h_0(x).
\end{align*}

%-------------------
%-------------------
%-------------------
\section{Details regarding the Robot Failures Case Study}\label{sec:robotfailures_details}

In Section 5 of the main article, we apply the proposed control chart and the PCA based control chart on a dataset from \citet{Dua2019, Camarinha-Matos1996} which contain force and torque measurements of a robot arm while executing the tasks of grasping an object, transferring it to a new location and ungrasping it.

\begin{figure}
    \centering
        \subfloat[Assembly cell integrating infrastructure (B-LEARN project)][Assembly cell integrating infrastructure (B-LEARN project)]{
        \includegraphics[width=0.5\linewidth]{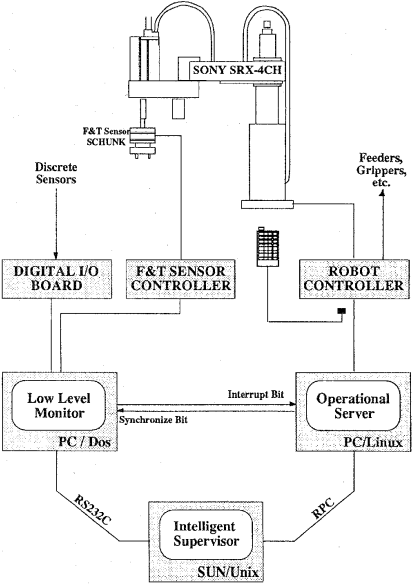}
        \label{fig:assembly infrastructure}
        }
        % \quad
        \subfloat[Regions of a gripper that may be involved in a failure][Regions of a gripper that may be involved in a failure]{
        \includegraphics[width=0.5\textwidth]{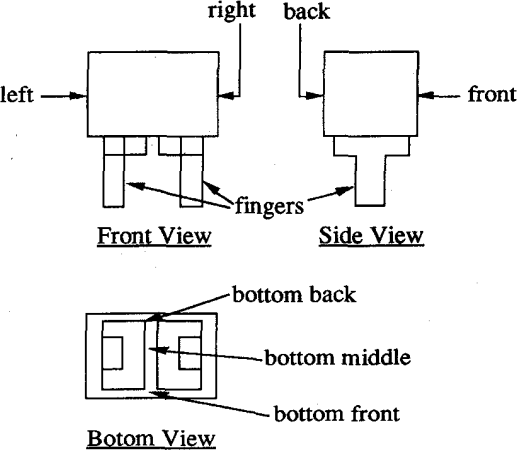}
        \label{fig:visRFcorr2}
        }
    \caption{\newtextforblind{\blind}{Visualizations of the Robot Arm taken from \cite{Camarinha-Matos1996}.}}
    \label{fig:Camarinha-Matos}
\end{figure}

\begin{figure}
        \centering
        \includegraphics[width = 0.91\linewidth]{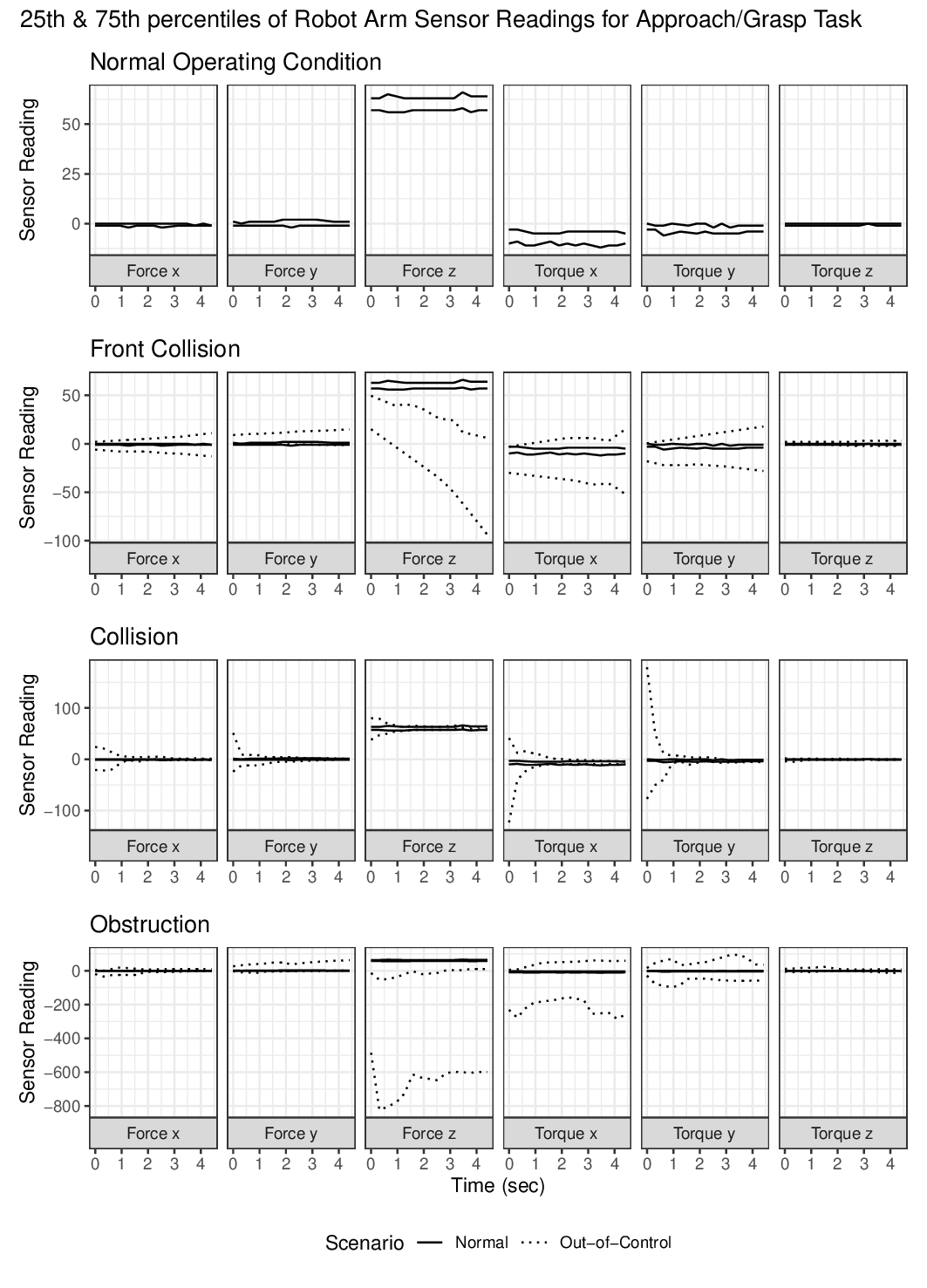}
        \caption{\newtextforblind{\blind}{A comparison of the sensor readings between in-control and out-of-control robot arm operation.  
        For each out-of-control scenario, we plot fitted percentiles of the observed sensor readings across time of execution.}}
        \label{fig:Robot Failures}
    \end{figure}

\newtextforblind{\blind}{Visualizations of the robot arm taken from \cite{Camarinha-Matos1996} are provided in Figure \ref{fig:Camarinha-Matos}}. 
Force and torque measurements are collected along three axes every 315 ms for 4.725s under various task completion and failure scenarios. 
Table \ref{tab:RobotFailuresSummary} provides a summary of the types of task completion and failure scenarios. 
\newtextforblind{\blind}{Figure \ref{fig:Robot Failures} provides a visualization of the various failure scenarios in comparison the the normal operating scenarios for the LP-1 Task where a robot arm approaches and grasps an object. }

\begin{table}[ht!]
\centering
\scalebox{0.8}{
\begin{tabular}{cccc}
Task ID &
  Failures in $\dots$ &
  Class &
  \begin{tabular}[c]{@{}c@{}}Sample\\ Size\end{tabular} \\ \hline \hline
LP-1 &
  approach-grasp &
  \begin{tabular}[c]{@{}c@{}}normal\\ collision\\ front collision\\ obstruction\end{tabular} &
  \begin{tabular}[c]{@{}c@{}}21\\ 17\\ 16\\ 34\end{tabular} \\ \hline
LP-2 &
  \begin{tabular}[c]{@{}c@{}}transfer of a part\\ (system behavior)\end{tabular} &
  \begin{tabular}[c]{@{}c@{}}normal\\ front collision\\ back collision\\ collision to the right\\ collision to the left\end{tabular} &
  \begin{tabular}[c]{@{}c@{}}20\\ 6\\ 7\\ 5\\ 9\end{tabular} \\ \hline
LP-3 &
  \begin{tabular}[c]{@{}c@{}}transferred location\\ (handled part)\end{tabular} &
  \begin{tabular}[c]{@{}c@{}}ok\\ slightly moved\\ moved\\ lost\end{tabular} &
  \begin{tabular}[c]{@{}c@{}}20\\ 15\\ 9\\ 3\end{tabular} \\ \hline
LP-4 &
  approach-ungrasp &
  \begin{tabular}[c]{@{}c@{}}normal\\ collision\\ obstruction\end{tabular} &
  \begin{tabular}[c]{@{}c@{}}24\\ 72\\ 21\end{tabular} \\ \hline
LP-5 &
  motion with part &
  \begin{tabular}[c]{@{}c@{}}normal\\ bottom collision\\ bottom obstruction\\ collision in part\\ collision in tool\end{tabular} &
  \begin{tabular}[c]{@{}c@{}}44\\ 26\\ 21\\ 47\\ 26\end{tabular} \\\hline
\end{tabular}
}
\caption{A summary of the tasks in the Robot Failures dataset from \citep{Camarinha-Matos1996} which can be found on \cite{Dua2019}.
}
\label{tab:RobotFailuresSummary}
\end{table}

We use the data from the LP-1 Task ID where the robot arm approaches and grasps an object as all other tasks seemed to have multiple ways to be IC. 
Of the 21 normal profiles in LP-1 we remove three profiles as their residuals after mean (profile) centering are nearly perfectly correlated.

\newtextforblind{\blind}{Observe in Figure \ref{fig:Robot Failures}, if the six sensor readings for a single profile across the 4.725s window are stacked in a single vector, the amount of variability of the in-control function is quite large especially in comparison to the variability around the in-control function.
That is to say, for this dataset $\Var[f] >> \sigma^2$.
Despite the ``small'' sample size of $n=90$ for each profile in this dataset and the small values of $m$ and $w$ used in the simulation study, the FAR of the EPCC is still quite low due to the large value of $\Var[f] / \sigma^2$.}

\begin{figure}[]
        \centering
        \subfloat[][]{
        \includegraphics[width=0.5\textwidth]{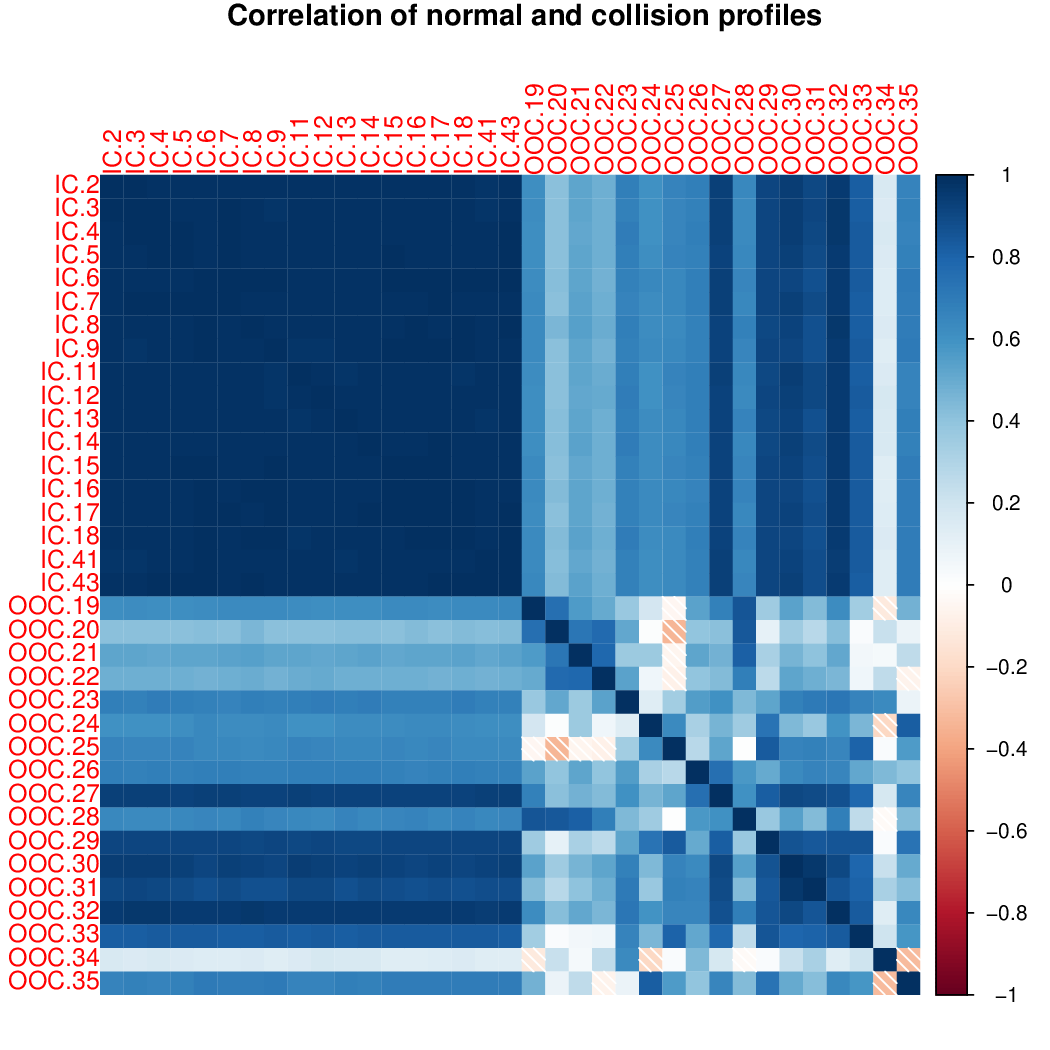}
        \label{fig:visRFcorr1}
        }
        % \quad
        \subfloat[][]{
        \includegraphics[width=0.5\textwidth]{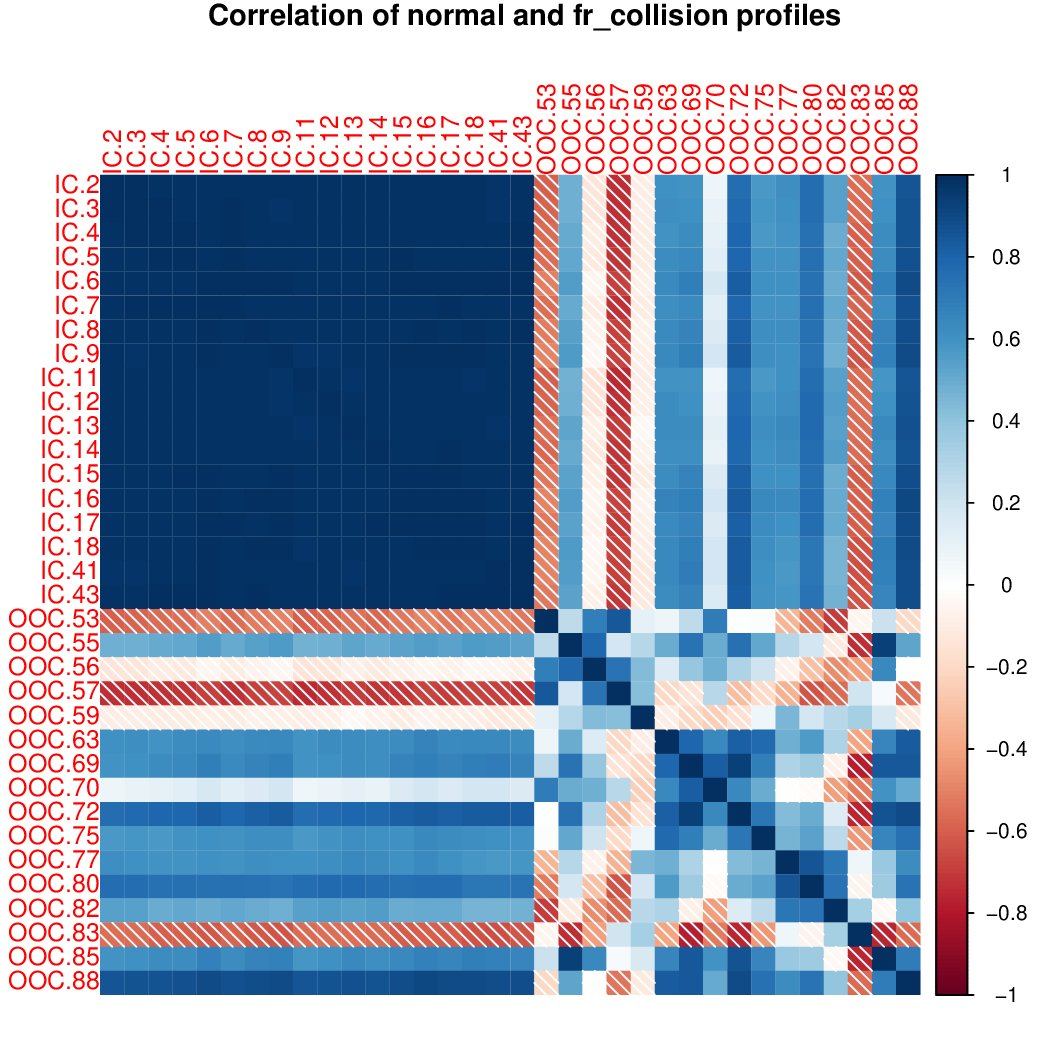}
        \label{fig:visRFcorr2}
        }
        \quad
        \subfloat[][]{
        \includegraphics[width=0.5\textwidth]{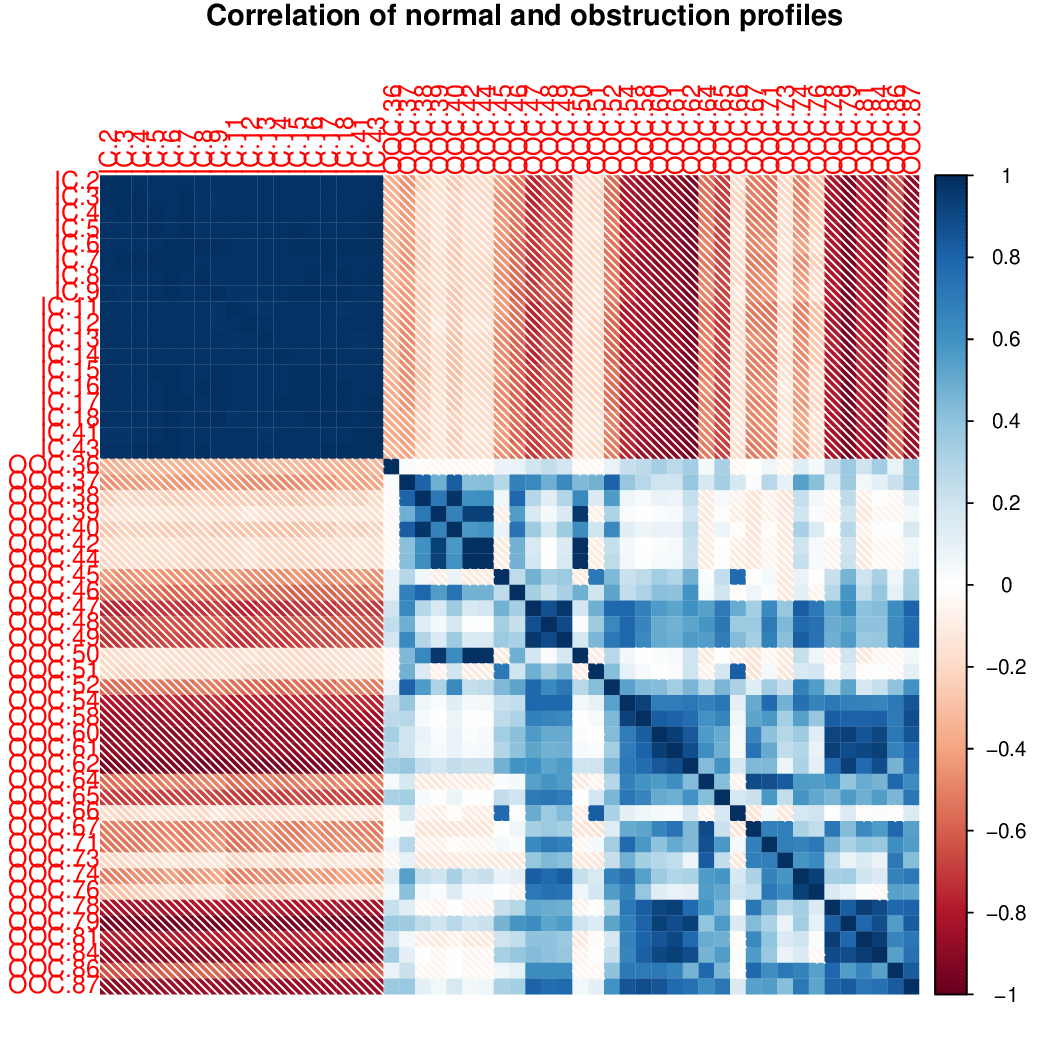}
        \label{fig:visRFcorr3}
        }
        \caption[]{\newtextforblind{\blind}{The sample correlation between profiles in the Robot Failures Dataset.
        Rows/columns labeled with an `IC' (`OOC') are the in-control (out-of-control) profiles where sensor readings are taken under normal (abnormal) operating conditions.}}
        \label{fig:RobotFailureCorr}
    \end{figure}

\newtextforblind{\blind}{
Regarding the control chart's out-of-control performance, notice the correlation structure between the IC and OOC profiles in   Figure \ref{fig:RobotFailureCorr} has a clear 2 $\times$ 2 block structure where the upper left block is the correlation of two IC profiles.
The off diagonal blocks are correlations of an IC profile with an OOC profile, and the lower right block corresponds to the correlation between two OOC profiles.
Compare Figure \ref{fig:RobotFailureCorr} in this response to Figure 1 in the main manuscript and observe the correlation matrices in Figure \ref{fig:RobotFailureCorr} deviates further from Figure 1a from the main manuscript than Figure 1b from the main manuscript does in all but the upper left block.  
This should suggest that our control chart should have good out-of-control performance, which is supported by the simulation study in the main manuscript. 
}

\newpage

\bibliographystyle{Chicago}
% \bibliography{references}